\documentclass[manuscript,screen]{acmart}
\AtBeginDocument{%
  }

\setcopyright{acmlicensed}
\copyrightyear{2018}
\acmYear{2018}
\acmDOI{XXXXXXX.XXXXXXX}
\acmConference[Conference acronym 'XX]{Make sure to enter the correct
  conference title from your rights confirmation email}{June 03--05,
  2018}{Woodstock, NY}
\acmISBN{978-1-4503-XXXX-X/2018/06}

\newcommand{\cheng}{}

\usepackage{tcolorbox}
\usepackage{multirow}
\usepackage[normalem]{ulem}
\usepackage{graphicx}
\usepackage{subfig}
\usepackage{enumitem}
\useunder{\uline}{\ul}{}

\begin{document}

\title{Facet-Aware Multi-Head Mixture-of-Experts Model with Text-Enhanced Pre-training for Sequential Recommendation}


\author{Mingrui Liu}
\affiliation{%
  \institution{Nanyang Technological University}
  \city{Singapore}
  \country{Singapore}}
\email{mingrui001@e.ntu.edu.sg}

\author{Sixiao Zhang}
\affiliation{%
  \institution{Nanyang Technological University}
  \city{Singapore}
  \country{Singapore}}
\email{sixiao001@e.ntu.edu.sg}

\author{Cheng Long}
\affiliation{%
  \institution{Nanyang Technological University}
  \city{Singapore}
  \country{Singapore}}
\email{c.long@ntu.edu.sg}

\renewcommand{\shortauthors}{Trovato et al.}

\begin{abstract}

Sequential recommendation (SR) systems excel at capturing users' dynamic preferences by leveraging their interaction histories. Most existing SR systems assign a single embedding vector to each item to represent its features, adopting various models to combine these embeddings into a sequence representation that captures user intent. However, we argue that this representation alone is insufficient to capture an item's multi-faceted nature (e.g., movie genres, starring actors). Furthermore, users often exhibit complex and varied preferences within these facets (e.g., liking both action and musical films within the genre facet), which are challenging to fully represent with static identifiers.
To address these issues, we propose a novel architecture titled \textit{\textbf{\underline{F}}acet-\textbf{\underline{A}}ware \textbf{\underline{M}}ulti-Head Mixture-of-\textbf{\underline{E}}xperts Model for Sequential Recommendation} (\textbf{\textit{FAME}}). We leverage sub-embeddings from each head in the final multi-head attention layer to predict the next item separately, effectively capturing distinct item facets. A gating mechanism then integrates these predictions by dynamically determining their importance. Additionally, we introduce a Mixture-of-Experts (MoE) network within each attention head to disentangle varied user preferences within each facet, utilizing a learnable router network to aggregate expert outputs based on context.
Complementing this architecture, we design a \textbf{Text-Enhanced Facet-Aware Pre-training} module to overcome the limitations of randomly initialized embeddings. By utilizing a pre-trained text encoder and employing an alternating supervised contrastive learning objective, we explicitly disentangle facet-specific features from textual metadata (e.g., descriptions) before sequential training begins. This ensures that the item embeddings are semantically robust and aligned with the downstream multi-facet framework. Extensive experiments on four public datasets demonstrate that our proposed method significantly outperforms existing baselines, validating the effectiveness of both the FAME architecture and the text-enhanced pre-training strategy.

\end{abstract}

\ccsdesc[500]{Information systems~Recommender systems}

\keywords{Recommender System, Sequential Recommendation}


\maketitle

\section{Introduction}

The exponential growth of online information presents users with a vast and ever-expanding sea of items, ranging from e-commerce products~\cite{product} and digital apps~\cite{content} to streaming videos~\cite{video1, video2}. With limited time to explore this abundance, recommender systems (RS) have become indispensable tools for facilitating efficient and satisfying decision-making. However, user interests are inherently dynamic, evolving continuously over time, which challenges platforms to deliver consistently relevant recommendations~\cite{wang2021survey}. To address this, sequential recommendation (SR) has emerged as a predominant paradigm. By leveraging the chronological sequence of user interactions, SR models capture the evolution of user preferences to predict future actions~\cite{fang2020deep, wang2019sequential}.

\begin{figure}[htbp]
  \centering
  \includegraphics[width=0.82\linewidth]{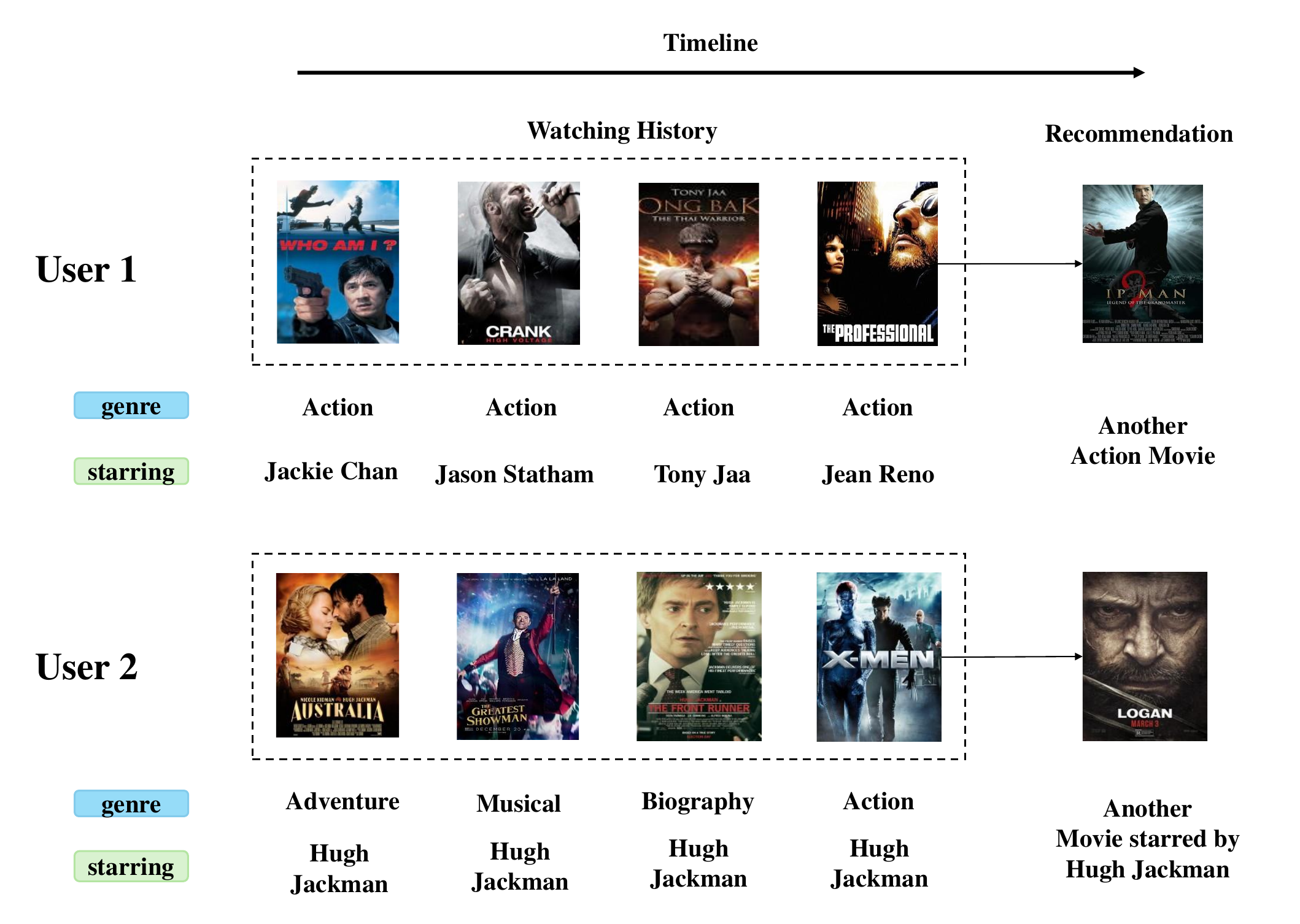}
  \caption{A motivating example illustrating multi-faceted user intent.}
  \label{fig: example}
  \Description{A motivation example}
\end{figure}

Traditional SR frameworks typically assign a single static embedding vector to represent each item. Architectures such as Recurrent Neural Networks (RNNs)~\cite{GRU4Rec, hidasi2015session}, attention-based models~\cite{SASRec, BERT4Rec, FDSA, Autoint}, and graph-based models~\cite{SRGNN, SURGE, DCRec, MSGIFSR} aggregate these item embeddings into a sequence representation to model user intent. The prediction is then usually performed by selecting the item with the highest compatibility (e.g., inner product) with this sequence vector. However, a monolithic embedding fails to adequately capture the multi-faceted nature of items (e.g., a movie possessing both \textit{Genre} and \textit{Starring} attributes)~\cite{Re4, MiasRec}. This limitation is critical when different facets drive user intent in different contexts.
As illustrated in Figure~\ref{fig: example}, User 1's history indicates a strong preference for the \textit{Action} genre; thus, recommending another action movie is appropriate. Conversely, User 2's interactions span multiple genres but strictly feature the actor Hugh Jackman, suggesting that the \textit{Starring} facet dominates their decision-making. These examples underscore that user interest is often driven by specific, distinguishable facets. Furthermore, in realistic scenarios, users may exhibit multiple distinct preferences within a single facet. For instance, within the \textit{Genre} facet, a user might enjoy both Action and Musical movies. Failing to disentangle the dominant facet and the specific preferences within it leads to suboptimal representations and recommendations.

Existing research has attempted to address intent complexity by using hierarchical windows~\cite{MSGIFSR, atten-mixer} or retrieving multi-item representations~\cite{MiasRec}. However, these methods largely focus on aggregating historical interactions and still neglect the inherent multi-faceted semantics of the items themselves.

To address these limitations, we propose a novel architecture titled \textit{\textbf{\underline{F}}acet-\textbf{\underline{A}}ware \textbf{\underline{M}}ulti-Head Mixture-of-\textbf{\underline{E}}xperts Model for Sequential Recommendation} (\textbf{\textit{FAME}}). We repurpose the multi-head attention mechanism to serve as a multi-facet predictor: the sub-embeddings from each head are utilized to independently predict the next item, with each head specializing in a distinct item facet. A gating mechanism then dynamically integrates these predictions based on their contextual importance. Furthermore, to disentangle complex user preferences within each facet, we replace the standard query projection with a Mixture-of-Experts (MoE) network. A learnable router assigns weights to different experts, effectively identifying the user's specific sub-preference (e.g., "Action" vs. "Musical") within the active facet.

Complementing the FAME architecture, we introduce a \textbf{Text-Enhanced Facet-Aware Pre-training} module to address the initialization bottleneck. Standard ID-based embeddings suffer from the cold-start problem and lack semantic interpretability. To bridge this gap, we leverage a pre-trained text encoder (e.g., BERT~\cite{BERT}) to extract rich semantic features from item metadata (e.g., product descriptions or movie plots). However, raw textual embeddings often represent semantics in an entangled manner. To explicitly align these features with our multi-facet framework, we project the initial text embeddings into $H$ disjoint subspaces using independent experts (MLP layers), where $H$ matches the number of attention heads in FAME. Within each subspace, we employ an alternating supervised contrastive learning objective. This explicitly pulls items sharing the same facet label (e.g., same Genre) closer while pushing others apart, thereby ensuring that the initialized embeddings are already disentangled and facet-aware before the sequential training begins. We refer to this extended framework with text-enhanced initialization as \textbf{FAME+}.

To summarize, this article extends our preliminary work~\cite{FAME} by incorporating semantic initialization and conducting a more comprehensive analysis. Our main contributions are as follows:
\begin{itemize}
    \item \textbf{(Presented in~\cite{FAME})} We propose a Multi-Head Prediction Mechanism that reinterprets attention heads as facet-specific predictors, enabling the model to capture diverse item attributes without increasing parameter complexity.
    \item \textbf{(Presented in~\cite{FAME})} We introduce a Mixture-of-Experts (MoE) network within the self-attention layer to disentangle multiple fine-grained user preferences within each facet, seamlessly integrating with existing transformer-based backbones.
    \item \textbf{(New in this work)} We design a text-enhanced pre-training framework that utilizes supervised contrastive learning to explicitly extract and disentangle facet-specific features from textual metadata, providing a robust initialization for the downstream model.
    \item \textbf{(Extended evaluation)} We conduct extensive experiments on four public datasets, demonstrating significant effectiveness compared to various baseline categories (sequential, pre-trained, multi-intent) on four public datasets.
\end{itemize}

\section{Related Work}
\subsection{Sequential Recommendation}
Recent advancements in neural networks and deep learning have spurred the development of various models to extract rich latent semantics from user behavior sequences and generate accurate recommendations. Convolutional Neural Networks (CNNs)~\cite{Caser}, Recurrent Neural Networks (RNNs)~\cite{GRU4Rec}, Transformer-based models~\cite{SASRec, BERT4Rec, FDSA, Autoint}, and Graph Neural Networks (GNNs)~\cite{SRGNN, SURGE, DCRec, MSGIFSR} have been widely employed to enhance representation learning and recommendation performance.
Self-supervised learning (SSL) has emerged as a promising technique for sequential recommendation~\cite{CL4SRec, ICLRec, Re4, DuoRec, DCRec, SSLRec}, with methods like CL4SRec~\cite{CL4SRec} and ICLRec~\cite{ICLRec} employing data augmentation and contrastive learning to improve sequence representations and capture user intents. Additionally, research has focused on modeling multiple user intents, such as the hierarchical window approach in MSGIFSR~\cite{MSGIFSR} and Atten-Mixer~\cite{atten-mixer}, or the multi-item-based representation in MiasRec~\cite{MiasRec}.
Furthermore, incorporating auxiliary information like item categories or attributes~\cite{cai2021category, S3-Rec} and textual descriptions~\cite{rajput2024recommender} has been explored to enrich item representations. The integration of large language models (LLMs) is another emerging trend in the field~\cite{LLM1, LLM2, llamarec}. However, these approaches are beyond the scope of this paper.

\subsection{Sparse Mixtures of Experts (SMoE)}
The Mixture-of-Experts (MoE) architecture has emerged as a powerful tool for handling complex tasks by distributing computations across multiple specialized models, or experts. While MoE models can significantly enhance model capacity, their computational overhead due to routing data to all experts can be prohibitive. To address this, Sparse Mixture of Experts (SMoE) was introduced, enabling each data point to be processed by a carefully selected subset of experts~\cite{MOE1, MOE2, switchtrans}. This approach offers the potential for substantial computational savings without compromising performance.

While SMoE has shown promise in various domains, its application in sequential recommendation remains relatively under-explored. 
Leveraging SMoE in this context could unlock new opportunities to enhance recommendation quality by effectively capturing and modeling diverse user preferences within a sequence.

\section{Preliminaries}

\subsection{Notations and Problem Statement}
Let $\mathcal{U}$ and $\mathcal{V}$ represent the user set and item set, where $u\in \mathcal{U}$ (resp. $v\in \mathcal{V}$) denotes an individual user (resp. item). Consequently, $|\mathcal{U}|$ and $|\mathcal{V}|$ denote the {\cheng sizes} of user set and item set, {\cheng respectively}. For each user $u$, we define their interaction sequence $\mathcal{S}_u = \{ \mathbf{v}_{1}^{(u)}, \cdots, \mathbf{v}_{i}^{(u)}, \cdots, \mathbf{v}_{t}^{(u)}\}$ as a chronologically ordered list of items. Here, $\mathbf{v}_{i}^{(u)}\in \mathcal{V}$ represents the item that user $u$ interacted with at time step $i$, and $t$ denotes the length of the interaction sequence for user $u$. 
Given a user interaction sequence $\mathcal{S}_u$, the goal of sequential recommendation is to predict the item that user $u$ will interact with at the next time step, $t+1$. Formally, we can define the problem as:
\begin{equation}
    \mathbf{v}_{u}^{(*)}=\mathop{\arg\max}\limits_{\mathbf{v}_i \in \mathcal{V}} P(\mathbf{v}_{t+1}^{(u)}=\mathbf{v}_i \mid \mathcal{S}_u)
\end{equation}

\subsection{Multi-Head Self-Attention}
\label{multiheadselfatt}
\begin{enumerate}
    \item \textbf{Item Embeddings:} The model first obtains embeddings for each item in the sequence (denoted as $x\in \mathbb{R}^{d}$).
    \item \textbf{Query, Key, Value Vectors:} Each item embedding ($x$) is then projected into three vectors:
    \begin{itemize}
        \item Query Vector ($q$): Represents what the model is currently looking for in the sequence.
        \item Key Vector ($k$): Captures the content of the current item.
        \item Value Vector ($v$): Contains the actual information associated with the item.
    \end{itemize}
    These projections are calculated using three trainable weight matrices (denoted by $\mathbf{W}_Q, \mathbf{W}_K\in \mathbb{R}^{d\times d_{k}}, \mathbf{W}_V\in \mathbb{R}^{d\times d_{v}}$):
    \begin{equation}
            \label{eq: KQV}
            \mathbf{q}=\mathbf{x}^T\cdot \mathbf{W}_Q,\quad \mathbf{k}=\mathbf{x}^T\cdot \mathbf{W}_K,\quad \mathbf{v}=\mathbf{x}^T\cdot \mathbf{W}_V,
    \end{equation}
    where $d_{k}$ is the dimension of query and key vector, and $d_{v}$ is the dimension of value vector.
    \item \textbf{Attention Scores:} The model calculates an attention score ($\alpha_{ij}$) for each pair of items $(i, j)$ in the sequence. This score reflects the similarity between the current item's query vector ($\mathbf{q}_i$) and the key vector ($\mathbf{k}_j$) of each other item. A normalization term ($\sqrt{d}$) is used to account for the vector dimension. The attention scores are then normalized using a softmax function (denoted by $\tilde{\alpha_{ij}}$) to create a probability distribution across all items, indicating the relative importance of each item to the current one.
    \begin{equation}
        \alpha_{ij} = \frac{\mathbf{q}_{i}^{T}\cdot \mathbf{k}_{j}}{\sqrt{d}},\quad \tilde{\alpha_{ij}}=\frac{\exp(\alpha_{ij})}{\sum_{j=1}^{t}\exp(\alpha_{ij})}
    \end{equation}
    \item \textbf{Item representation:} {\cheng The} item representation $\mathbf{f}_i$ is calculated based on weighted sum of value vectors in the sequence. The weights for this summation are derived from the previously calculated attention scores ($\tilde{\alpha_{ij}}$):
    \begin{equation}
        \mathbf{f}_i = \sum_{j=1}^{t}\tilde{\alpha_{ij}}\cdot \mathbf{v}_j
        \label{eq: f}
    \end{equation}
\end{enumerate}

\section{Methods}
\label{Methods}

\subsection{Overview}
\begin{figure*}[h]
  \centering
  \includegraphics[width=\linewidth]{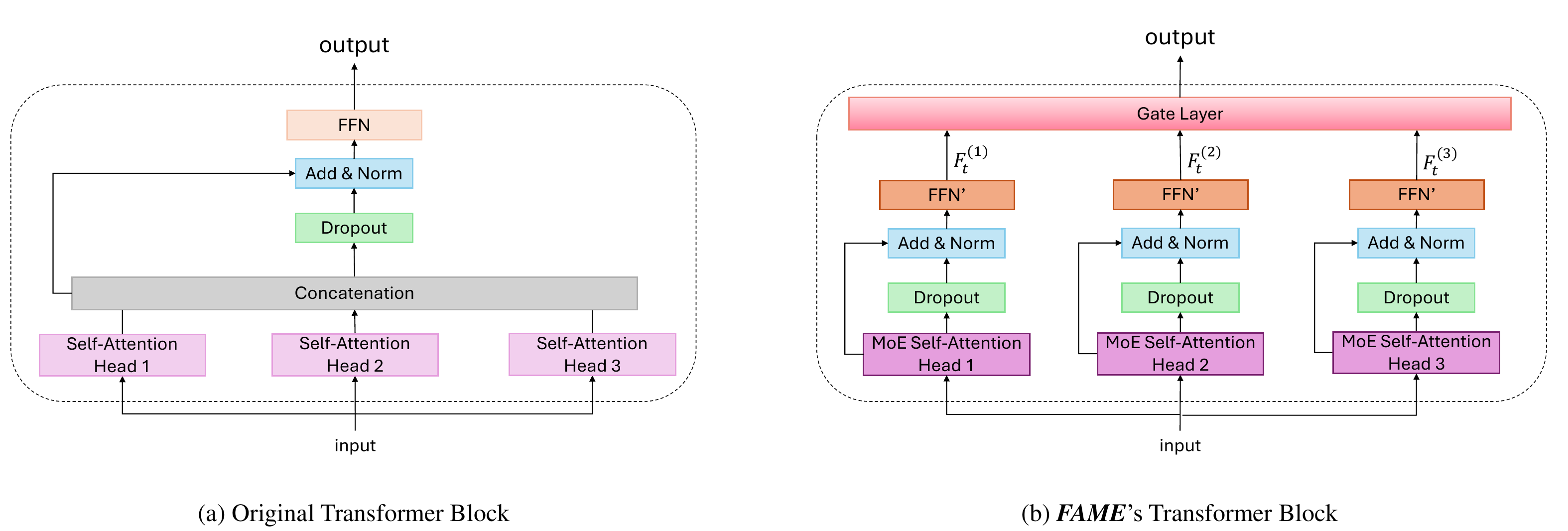}
  \caption{Overview of the proposed model: {\cheng (a)} illustrates the original Transformer block, while {\cheng (b)} depicts the architecture of our proposed \textit{\textbf{FAME}} model. For simplicity, the LayerNorm and Dropout operations following the FFN (FFN') are omitted from the Figure}
  \label{fig: overview}
  \Description{The overview of the model}
\end{figure*}

This section introduces our proposed framework with a high-level overview, which is displayed in Figure~\ref{fig: overview}. 
The framework incorporates two key components: the \textbf{Facet-Aware Multi-Head Prediction Mechanism} (detailed in Section.~\ref{MultiHeadPLayer}), which learns to represent each item with multiple sub-embedding vectors, each capturing a specific facet of the item; and the \textbf{Mixture-of-Experts Self-Attention Layer} (detailed in Section.~\ref{MOEAttLayer}), which employs a Mixture-of-Experts (MoE) network within each subspace to capture the users' specific preferences within each facet.
{\cheng Our framework can be seamlessly integrated to any attention-based recommendation model. In this paper, we incorporate our framework to SASRec for illustration.}

\subsection{Facet-Aware Multi-Head Prediction Mechanism}
\label{MultiHeadPLayer}
\subsubsection{Original SASRec Prediction Process}
\label{SASRec}
In the original SASRec model, the final prediction for the next item is based on the last item's representation ($\mathbf{f}_{t}$, calculated by Equation~\ref{eq: f}, which can also be regarded as the sequence representation) obtained from the last self-attention layer. This representation is processed through a feed-forward network (FFN) with ReLU activation for non-linearity, followed by layer normalization, dropout, and a residual connection:
\begin{equation}
    \begin{split}
        {\rm FFN}(\mathbf{f}_{t})={\rm RELU}(\mathbf{f}_{t}^{T}\cdot \mathbf{W}_1 + b_1)^{T}\cdot \mathbf{W}_2 + b_2,\\
        \mathbf{F}_t={\rm LayerNorm}(\mathbf{f}_{t} + {\rm Dropout}({\rm FFN}(\mathbf{f}_{t}))),
    \end{split}
    \label{eq: FFN}
\end{equation}
Here, $\mathbf{W}_1, \mathbf{W}_2\in \mathbb{R}^{d\times d}, b_1, b_2\in \mathbb{R}^{d}$ are all learnable parameters.
The final user preference score for item $v$ at step $(t+1)$ is then calculated as the dot product between the item embedding ($\mathbf{x}_v$) and the sequence representation ($\mathbf{f}_t$):
\begin{equation}
    P(\mathbf{v}_{t+1}=v|\mathcal{S}_u)=\mathbf{x}_{v}^{T}\cdot \mathbf{F}_t,
\end{equation}
Top-$k$ items with the highest preference scores are recommended to the user.

\subsubsection{Motivation for Our Approach}
\label{multiheadmotivation}

The multi-head self-attention mechanism splits the sequence representation and item embeddings into multiple subspaces (heads). Research suggests that these heads can allocate different attention distributions so as to perform different tasks~\cite{transformer}. We hypothesize that these heads could also capture different facets of items (e.g., genre and starring actors in the context of movie recommendation). This ability to capture multi-faceted information has the potential to improve recommendation quality.

\subsubsection{Proposed Multi-Head Recommendation}
Instead of performing a single attention function with $d$-dimensional keys, values and queries, it is found beneficial to linearly project the queries, keys and values $H$ 
times with different, learned linear projections to $d_{k}$, $d_{k}$ and $d_{v}$ dimensions, respectively~\cite{transformer}.
{\cheng Here, $H$ is the number of heads, and 
$d_{k}$, $d_{k}$ and $d_{v}$ are typically set to $d'=\frac{d}{H}$.}

Leveraging the multi-head attention mechanism, we propose a novel approach where each head independently generates recommendations. The final item embedding from head $h$ is denoted as $\mathbf{f}_{t}^{(h)}\in \mathbb{R}^{d'}$. We then process this embedding similarly {\cheng as we do for} the original model:
\begin{equation}
    \begin{gathered}
        {\rm FFN'}(\mathbf{f}_{t}^{(h)})={\rm RELU}(\mathbf{f}_{t}^{(h)T}\cdot \mathbf{W'}_1 + \mathbf{b'}_1)^{T}\cdot \mathbf{W'}_2 + \mathbf{b'}_2,\\
        \mathbf{F}_{t}^{(h)}={\rm LayerNorm}(\mathbf{f}_{t}^{(h)} + {\rm Dropout}({\rm FFN'}(\mathbf{f}_{t}^{(h)}))),
    \end{gathered}
    \label{eq: FFN'}
\end{equation}
Unlike the original FFN (Equation~\ref{eq: FFN}), the feed-forward network applied to each head (${\rm FFN'}$) operates on a reduced dimension of $d'$. The learnable parameters for ${\rm FFN'}$ are therefore adjusted accordingly: $\mathbf{W}'_1, W'_2\in \mathbb{R}^{d'\times d'}, b'_1, b'_2\in \mathbb{R}^{d'}$. This adaptation aligns with the dimensionality of sub-embeddings within each head.
To enhance parameter efficiency and improve performance, we adopt a shared feed-forward network (${\rm FFN'}$) across all attention heads.
Each head generates the preference score for each item independently, i.e., 
\begin{equation}
    P^{(h)}(\mathbf{v}_{t+1}=v|\mathcal{S}_u)=\mathbf{x}_{v}^{(h)T}\cdot \mathbf{F}^{(h)}_{t},
\end{equation}
where $\mathbf{x}_{v}^{(h)}\in \mathbb{R}^{d'} $ is the sub-embedding of the item $v$, reflecting the features of the specific facet corresponding to the attention head $h$. Specifically, it is calculated by a linear transformation from its original embedding:
\begin{equation}
    \mathbf{x}_{v}^{(h)} = \mathbf{x}_{v}^{T} \cdot \mathbf{W}_{f}^{(h)},
\end{equation}
with $\mathbf{W}_{f}^{(h)}\in \mathbb{R}^{d\times d'}$ being a learnable matrix.

In order to integrate the recommendation results from each head, we employ a gate mechanism to determine the relative importance of each head's recommendations:

\begin{equation}
    \begin{gathered}
        \mathbf{g}=\left[\mathbf{F}_{t}^{(1)}|\dots |\mathbf{F}_{t}^{(H)}\right]^T\cdot \mathbf{W}_g + \mathbf{b}_g, \\
            \tilde{\mathbf{g}}={\rm softmax}(\mathbf{g})
    \end{gathered}
    \label{eq: gate}
\end{equation}

Here, $\left[\cdot|\cdot\right]$ denotes the concatenation operation. Each element $\tilde{g}^{(h)}\in [0,1]$ within the vector $\tilde{g}$ represents the importance of head $h$ in determining the user's dominant interest or preference. For instance, a higher $\tilde{g}^{(h)}$ for a genre-focused head indicates a stronger preference for specific movie genres, while a higher value for an actor-focused head suggests a preference for movies starring particular actors. The gate mechanism, parameterized by $\mathbf{W}_g\in \mathbb{R}^{d\times H}$ and $b_g\in \mathbb{R}^{H}$, learns to assign appropriate weights to each head based on the user's current context.
Finally, we compute a unified preference score for each item by weighting the recommendations from each head:

\begin{equation}
    P(\mathbf{v}_{t+1}=\mathbf{v}|\mathcal{S}_u)=\sum_{i=1}^{H}\tilde{\mathbf{g}}^{(h)}\cdot P^{(h)}(\mathbf{v}_{t+1}=\mathbf{v}|\mathcal{S}_u)
\end{equation}

This approach allows the model to exploit the strengths of each head while assigning appropriate weights based on their importance in the specific context.

\subsection{Mixture-of-Experts Self-Attention Layer}
\label{MOEAttLayer}
\begin{figure}[htb]
  \centering
  \includegraphics[width=0.83\linewidth]{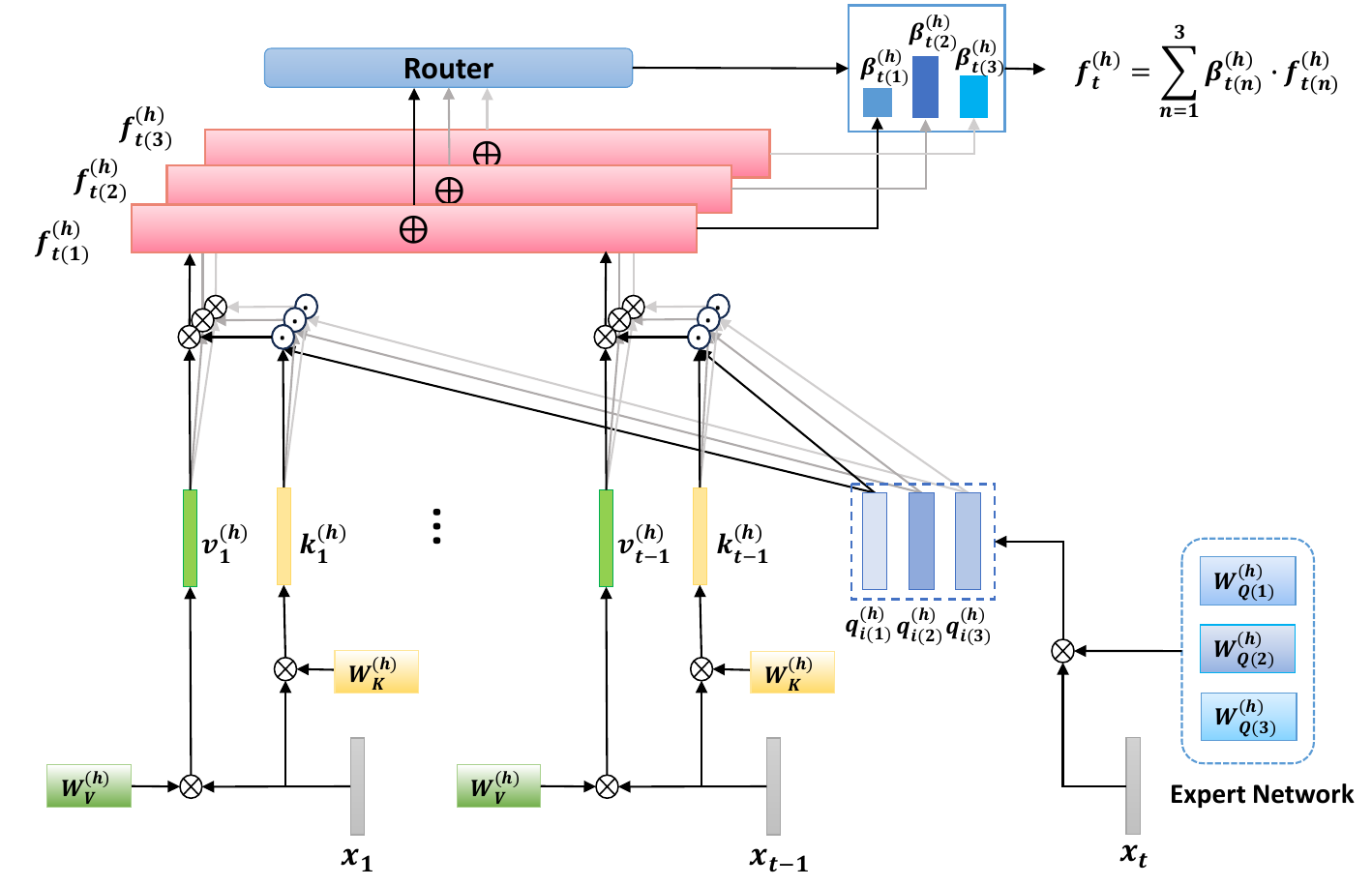}
  \caption{MoE Self-Attention Network: Integrated Item Representation Calculation.
This diagram visualizes the computational process for determining the integrated item representation of the final item ($\mathbf{f}_{t}^{(h)}$) within a specific head ($h$) of our proposed model.}
  \label{fig: MOE}
\end{figure}

While the Facet-Aware Multi-Head mechanism effectively captures item facets, users often exhibit more granular and diverse preferences within these facets. To address this, we introduce the Mixture-of-Experts Self-Attention Layer (MoE-Attention), as illustrated in Figure~\ref{fig: MOE}. 

We assume that each facet can be decomposed into $N$ distinct preferences. For instance, a genre facet might include preferences for action, comedy, musicals, etc.
To capture the nuanced preferences within each facet of a sequence, we replace the standard query generation mechanism in self-attention (Equation~\ref{eq: KQV}) with a Mixture-of-Experts (MoE) network in each head. This network consists of $N$ experts, each represented by a trainable matrix $\mathbf{W}_{Q(n)}^{(h)}\in \mathbb{R}^{d\times d'}$ (where $n\in [1,N]$). Each expert within a head is designed to capture one of these preferences by transforming an item embedding $\mathbf{x}_i$ (i.e., the embedding of the $i^{th}$ item in the sequence) into an expert query vector $\mathbf{q}_{i(n)}^{(h)}\in \mathbb{R}^{d'}$ as follows:
\begin{equation}
    \mathbf{q}_{i(n)}^{(h)}=\mathbf{x}_{i}^{T}\cdot \mathbf{W}_{Q(n)}^{(h)}.
    \label{eq: newq}
\end{equation}
The key vector ($\mathbf{k}_{j}^{(h)}$) and value vector ($\mathbf{v}_{j}^{(h)}$) of the $j^{th}$ sequence item in head $h$ are computed using the same linear transformations as in the original SASRec model:
\begin{equation}
    \mathbf{k}_{j}^{(h)} = \mathbf{x}_{j}^{T}\cdot \mathbf{W}_{K}^{(h)},\quad \mathbf{v}_{j}^{(h)} = \mathbf{x}_{j}^{T}\cdot \mathbf{W}_{V}^{(h)}.
\end{equation}
Then the attention score for the $i^{th}$ item relative to the $j^{th}$ item in head $h$ by expert $n$ is computed as:
\begin{equation}
    \begin{gathered}
        \mathbf{\alpha}_{ij(n)}^{(h)} = \frac{\mathbf{q}_{i(n)}^{(h)T}\cdot \mathbf{k}_{j}^{(h)T}}{\sqrt{\mathbf{d'}}},\\ \tilde{\alpha}_{ij(n)}^{(h)} = {\rm softmax}(\alpha_{i1(n)}^{(h)},\cdots, \alpha_{it(n)}^{(h)}).
    \end{gathered}
\end{equation}

\begin{figure}[htb]
  \centering
  \includegraphics[width=0.85\linewidth]{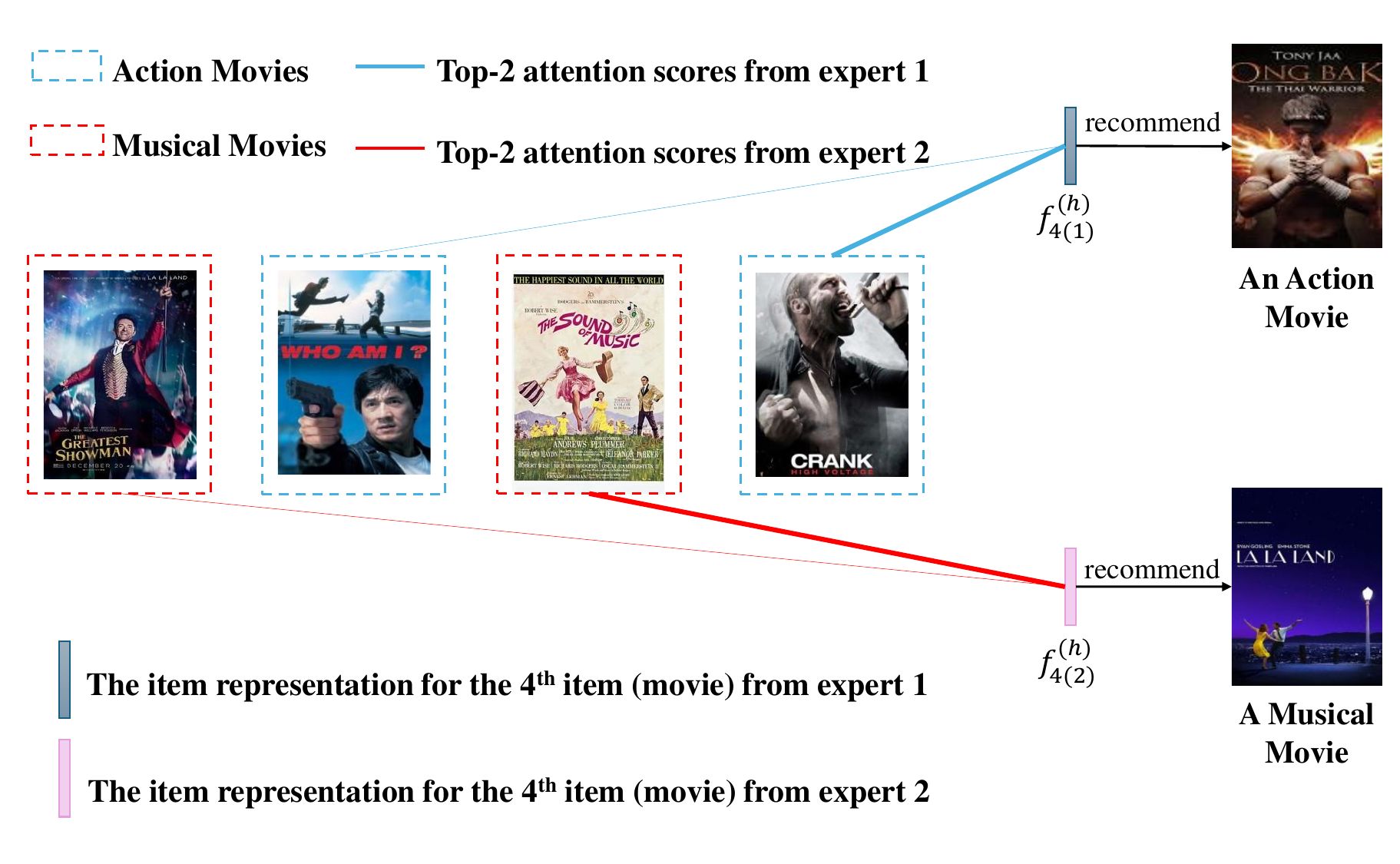}
  \caption{An example on attention scores distribution and recommendation results among different experts on genre-focused head}
  \label{fig: MOEexample}
\end{figure}

The item representation of the $i^{th}$ item for head $h$ and expert $n$ ($\mathbf{f}_{i(n)}^{(h)}$) is then calculated as a weighted sum of value vectors, where the weights are the corresponding attention scores:
\begin{equation}
    \mathbf{f}_{i(n)}^{(h)} = \sum_{j=1}^{t} \tilde{\mathbf{\alpha}}_{ij(n)}^{(h)}\cdot \mathbf{v}_{j}^{(h)}.
\end{equation}
As illustrated in Figure~\ref{fig: MOEexample}, consider a genre-focused head with two experts: one for action movies and another for musical movies. As detailed in Section.~\ref{SASRec}, the standard SASRec model treats the representation of the final item in a sequence as the overall sequence representation. 
To illustrate our MoE attention mechanism, we focus on the attention scores associated with the $4^{th}$ (and final) item in the sequence. The first expert's query vector of the $4^{th}$ item ($\mathbf{q}_{4(1)}^{(h)}$) would assign higher attention scores to action movies (items 1 and 3), while the second expert's query vector ($\mathbf{q}_{4(2)}^{(h)}$) would focus on musical movies (items 2 and 4). Consequently, the final item's representation ($\mathbf{f}_{4(1)}^{(h)}$) generated by the first expert would lean towards recommending action movies, whereas the representation ($\mathbf{f}_{4(2)}^{(h)}$) from the second expert would favor musical movies.

To dynamically determine the importance of each preference within a facet (e.g., whether action or musical is the preferred genre), we introduce a router network parameterized by $\mathbf{W}_{exp}^{(h)}\in \mathbb{R}^{(n\cdot d')\times n}$. This network assigns an importance score $\mathbf{\beta}_{i(n)}^{(h)}\in (0,1)$ to each item representation generated by each expert $\mathbf{f}_{i(n)}^{(h)}$.
The importance scores are computed as follows:
\begin{equation}
    \beta_{i\cdot}^{(h)} = softmax(\left[\mathbf{f}_{i(1)}^{(h)}|\cdots |\mathbf{f}_{i(n)}^{(h)}\right]^T\cdot \mathbf{W}_{exp}^{(h)}).
    \label{eq: expertimportance}
\end{equation}
The integrated item representation $\mathbf{f}_{i}^{(h)}$ for the $i^{th}$ item in head $h$ is then computed as a weighted sum of the expert query vectors:
\begin{equation}
    \mathbf{f}_{i}^{(h)} = \sum_{n=1}^{N}\beta_{i(n)}^{(h)}\cdot \mathbf{f}_{i(n)}^{(h)}.
    \label{eq: querysum}
\end{equation}
The integrated item representation ($\mathbf{f}_{i}^{(h)}$) represents the overall preference at the $i^{th}$ timestamp within head $h$. For instance, for the case in Figure~\ref{fig: MOEexample}, a higher weight for $\mathbf{f}_{4(1)}^{(h)}$ (resp. $\mathbf{f}_{4(2)}^{(h)}$) would {\cheng push} the model towards recommending action (resp. musical) movies.

\subsection{Deployment and Training}
\subsubsection{Model Deployment}
Our FAME model is built upon the SASRec (or any attention-based) framework, with the final Transformer layer replaced by our proposed architecture.
\subsubsection{Training Pipeline}
We initiate our model by pre-training an attention-based sequential recommendation model (e.g., SASRec). Subsequently, we replace Transformer block's query matrix at the final layer with our proposed MoE network ({\cheng Section}~\ref{MOEAttLayer}) while retaining the original key and value matrices. The newly introduced components, including the head-specific ${\rm FFN'}$ (Equation~\ref{eq: FFN'}), gate mechanism (Equation~\ref{eq: gate}), and router (Equation~\ref{eq: expertimportance}), are randomly initialized. The entire model is then fine-tuned end-to-end.

\subsubsection{Training Objectives}
A global cross-entropy loss function is employed to optimize the model during training:
\begin{equation}
    \label{eq: lossce}
    \mathcal{L}_{ce} = -\sum_{u\in \mathcal{U}}\log \left(\frac{\exp(\mathbf{x}_{t+1}^{(u)T}\cdot \mathbf{f}_{t}^{(u)})}{\sum_{i}\exp(\mathbf{x}_{i}^{T}\cdot \mathbf{f}_{t}^{(u)})} \right).
\end{equation}


\section{Text-Enhanced Facet-Aware Pre-training}
\label{sec:method_extension}

To overcome the inherent limitations of randomly initialized ID-based embeddings, namely, the cold-start problem and a lack of semantic interpretability, we propose a \textit{Text-Enhanced Facet-Aware Pre-training} framework. This module is designed to explicitly align item representations with distinct semantic attributes (e.g., Genre, Director) before they are ingested by the downstream FAME recommender, as detailed in Section~\ref{Methods}.

\begin{figure}[htb]
  \centering
  \includegraphics[width=0.82\linewidth]{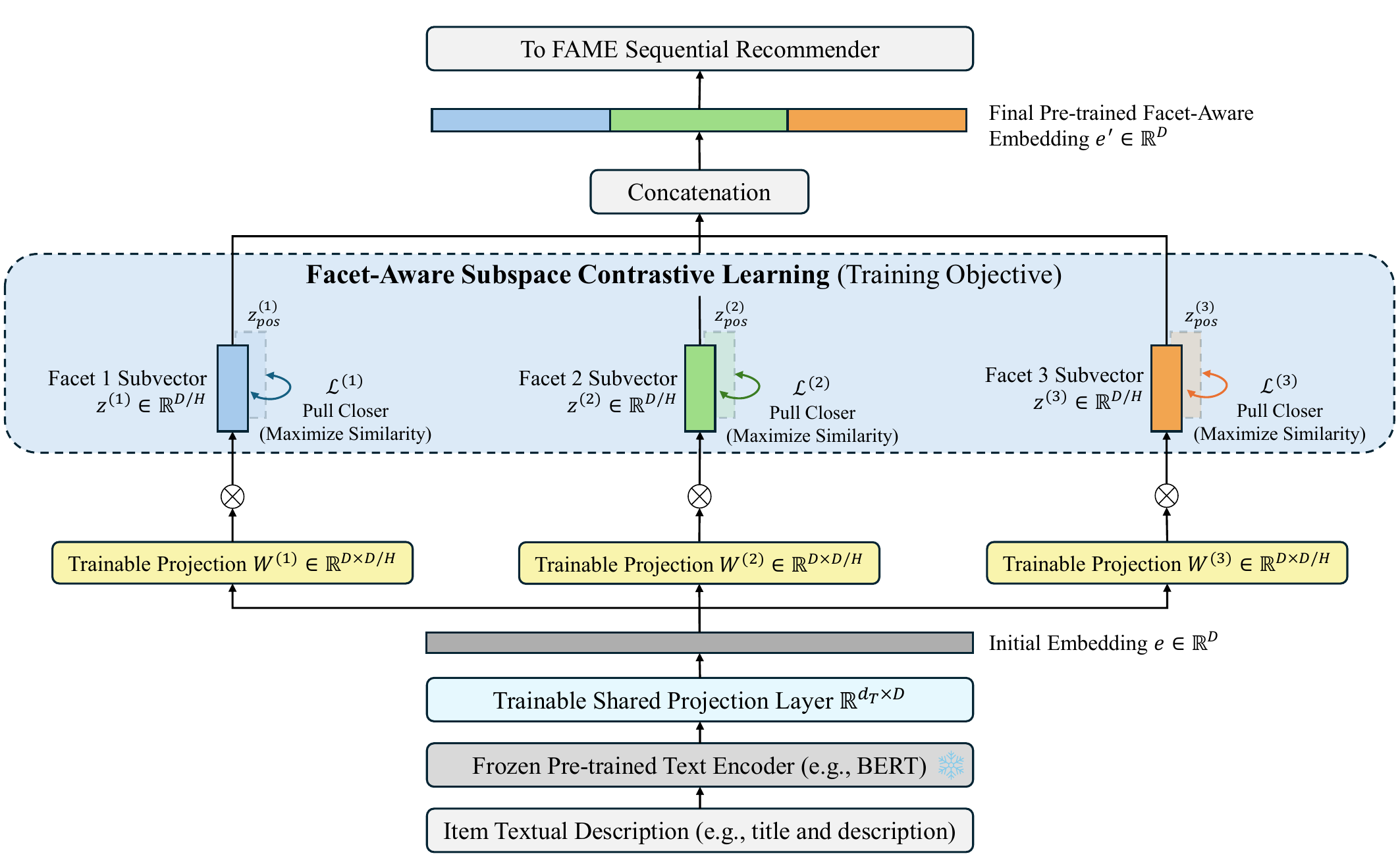}
  \caption{Overview of the Text-Enhanced Facet-Aware Pre-training Framework.}
  \label{fig: pretraining}
\end{figure}

An overview of the framework is presented in Figure~\ref{fig: pretraining}. The architecture comprises three integral components: (1) a multi-facet textual encoder that projects semantic information into disjoint subspaces; (2) an alternating supervised contrastive learning objective that effectively disentangles facet representations; and (3) a stratified sampling strategy to guarantee valid training signals within batches.

\subsection{Multi-Facet Textual Encoding}
Let $\mathcal{I}$ denote the set of items. For each item $i \in \mathcal{I}$, we utilize its textual metadata $\mathcal{T}_i$ (e.g., the title and description) as input. We employ a pre-trained language model (e.g., BERT~\cite{BERT}) as a frozen text encoder to extract a robust initial semantic vector:

\begin{equation}
    \mathbf{e}_i = \text{Encoder}(\mathcal{T}_i) \in \mathbb{R}^{D_{T}},
\end{equation}
where $\mathbf{e}_i$ represents the pooled output (e.g., the [CLS] token embedding).

To support the downstream FAME architecture, which relies on $H$ distinct experts to process different facets of an item, we project $\mathbf{e}_i$ into $H$ distinct subspaces. A shared Multi-Layer Perceptron (MLP) first transforms the text embedding into the recommender's latent space, followed by $H$ independent projection heads:

\begin{equation}
    \mathbf{h}_i = \sigma(\mathbf{W}_{shared} \cdot \mathbf{e}_i + \mathbf{b}_{shared}),
\end{equation}
\begin{equation}
    \mathbf{z}_i^{(h)} = \frac{\mathbf{W}^{(h)} \cdot \mathbf{h}_i + \mathbf{b}^{(h)}}{\| \mathbf{W}^{(h)} \cdot \mathbf{h}_i + \mathbf{b}^{(h)} \|_2}, \quad \forall h \in \{1, \dots, H\},
\end{equation}
where $\sigma$ denotes a non-linear activation function (e.g., ReLU), and $\mathbf{z}_i^{(h)} \in \mathbb{R}^{D/H}$ is the normalized subvector for the $h$-th facet. This normalization ensures that all embeddings reside on a hypersphere, stabilizing the subsequent contrastive learning.

\subsection{Alternating Supervised Contrastive Learning}
\label{subsec:supcon}

We adopt the Supervised Contrastive Learning (SupCon) framework~\cite{supervisedCL} to explicitly structure the embedding space. For a specific facet $h$, the objective is to minimize the distance between items sharing the same attribute label $y^{(h)}$ while maximizing the distance to items with different labels.

However, optimizing all facets simultaneously is infeasible due to the conflicting sampling requirements inherent in mini-batch training. Effective contrastive learning strictly requires that every mini-batch contains multiple positive samples (i.e., pairs sharing the same label) for the target facet. These distributions are often mutually exclusive; for instance, a mini-batch sampled to guarantee sufficient positive pairs for \textit{Genre} (e.g., containing 8 distinct 'Action' movies) effectively becomes a random sample with respect to \textit{Director}. In such a batch, it is statistically probable that all sampled movies feature different directors. Consequently, the \textit{Director} facet would suffer from a complete lack of positive pairs (zero valid signal), rendering the loss for that head uncomputable or noisy.

To resolve this conflict, we propose an \textbf{Alternating Optimization} strategy. Instead of a joint update, we decouple the training steps. In each iteration, we construct a mini-batch using the sampler specific to a single active head $h$ (guaranteeing positive pairs for that facet only). The training process cycles through facets $h=1 \dots H$, updating the shared parameters $\mathbf{W}_{shared}$ and the specific projection head $\mathbf{W}^{(h)}$ while keeping the other heads frozen.


The loss for a generic batch $\mathcal{B}$ on the active head $h$ is defined as:
\begin{equation}
\label{eq: pretraining_loss}
    \mathcal{L}^{(h)} = \sum_{i \in \mathcal{B}} \frac{-1}{|P(i)|} \sum_{p \in P(i)} \log \frac{\exp(\mathbf{z}_i^{(h)} \cdot \mathbf{z}_p^{(h)} / \tau)}{\sum_{a \in A(i)} \exp(\mathbf{z}_i^{(h)} \cdot \mathbf{z}_a^{(h)} / \tau)},
\end{equation}
where $A(i) \equiv \mathcal{B} \setminus \{i\}$ is the set of all indices excluding the anchor $i$, and $P(i) \equiv \{p \in A(i) \mid y_p^{(h)} = y_i^{(h)}\}$ is the set of distinct positive pairs in the batch. $\tau$ is a temperature hyperparameter.

\paragraph{Masking Strategy.} Efficient computation is achieved via a masking matrix $\mathbf{M} \in \{0, 1\}^{|\mathcal{B}| \times |\mathcal{B}|}$, where $M_{ij} = 1 (y_i^{(h)} = y_j^{(h)})$. Crucially, we enforce a zero diagonal ($M_{ii} = 0$) to strictly exclude self-contrast, ensuring the model learns solely from relationships between distinct items rather than trivial self-identities. The loss is computed by vectorizing the log-likelihood over the non-zero entries of $\mathbf{M}$.

\subsection{Stratified Facet Sampling}
\label{subsec:sampling}
While the objective in Equation~\ref{eq: pretraining_loss} is theoretically sound, applying it directly with standard random sampling presents a critical challenge: the \textbf{Label Sparsity Issue}. High-cardinality facets (e.g., \textit{Director} or \textit{Actor}) often possess thousands of unique classes. In a randomly sampled mini-batch, it is statistically probable that an item $i$ acts as the sole representative of its class $y_i^{(h)}$.

In such cases, the set of positive pairs $P(i)$ becomes empty ($|P(i)| = 0$). Crucially, our masking strategy explicitly enforces a zero diagonal ($M_{ii} = 0$), strictly prohibiting an item from serving as its own positive pair. This design is fundamental to our goal: we aim to learn semantic commonalities shared \textit{across distinct items} (e.g., capturing the shared "style" of two different Nolan movies) rather than optimizing for trivial self-identity. Consequently, if an item is an "orphan" in the batch, it lacks any valid positive partner to generate a "pull" signal, leading to null contributions to the loss and vanishing gradients.



To overcome this, we introduce a \textbf{Fair $P \times K$ Sampler}. This strategy guarantees valid positive pairs by constructing each mini-batch $\mathcal{B}^{(h)}$ to contain exactly $P$ distinct classes, with $K$ samples per class. Formally, the batch indices are generated such that:
\begin{equation}
    |\{ c \mid \exists i \in \mathcal{B}^{(h)}, y_i^{(h)} = c \}| = P, \quad \text{and} \quad \forall c, |\{ i \in \mathcal{B}^{(h)} \mid y_i^{(h)} = c \}| = K.
\end{equation}
This ensures that every item in the batch has exactly $K-1$ valid positive partners. Furthermore, we employ a "fair" epoch strategy that cycles through all valid classes (those containing at least $K$ samples) to ensure that long-tail items are represented equally during the pre-training phase.

\subsection{Integration with FAME}
Upon completion of the pre-training, we obtain subvectors $\mathbf{z}^{(h)}\in \mathbb{R}^{D/H}$ that capture the facet-specific characteristics of each item. These subvectors are concatenated to form the complete item embedding:
\begin{equation}
    \mathbf{e}_{i}'=\left[\mathbf{z}^{(1)} \mathbin{\|} \cdots \mathbin{\|} \mathbf{z}^{(H)}\right]\in \mathbb{R}^D.
\end{equation}
The resulting embedding $\mathbf{e}_{i}'$ encapsulates rich, facet-aware features and serves as the initialized input for the FAME framework.

\section{Experiments}

\subsection{Datasets}
\label{sec:datasets}

\subsubsection{Sequence Interaction Data}
\label{seqdata}
We conduct comprehensive experiments on four widely-used benchmark datasets. \textit{Beauty}, \textit{Sports}, and \textit{Toys} are subcategories from the Amazon Review dataset~\cite{Amazondata}, representing e-commerce user behaviors. \textit{ML-20m} is a prominent subset of the MovieLens dataset~\cite{movielens}, containing approximately 20 million movie ratings. 

Consistent with standard preprocessing protocols in sequential recommendation~\cite{S3-Rec, CL4SRec}, we apply a 5-core filter to all datasets, retaining only users and items with at least 5 interactions to ensure data quality. The detailed interaction statistics of the processed datasets are summarized in Table~\ref{tab: datasetstat}.

\subsubsection{Textual and Facet Metadata}
\label{sec:facet_data}
To support our Text-Enhanced Facet-Aware Pre-training module, we augment the interaction data with rich semantic metadata. For each item, we construct a raw textual input string for the initial encoder (e.g., BERT) and explicitly extract structured facet labels for the contrastive learning objectives.

\paragraph{Textual Input Construction.} 
We linearize the metadata into a structured string format to serve as the input for the frozen text encoder. The formats for the Amazon and MovieLens datasets are defined as follows:

\begin{tcolorbox}[colback=gray!10, colframe=gray!50, title=Textual Input Templates, arc=2mm]
\textbf{Amazon Datasets (Beauty, Sports, Toys):} \\
\texttt{"title": \{Title\}; "description": \{Desc\}; "category": \{Cat\}; "brand": \{Brand\}}

\vspace{0.5em}
\textbf{MovieLens-20m:} \\
\texttt{"title": \{Title\}; "description": \{Desc\}; "genres": \{Genres\}; "directors": \{Directors\}; "cast": \{Cast\}}
\end{tcolorbox}
\noindent Note that for ML-20m, metadata such as directors and cast were scraped from the official MovieLens website\footnote{\url{https://movielens.org/}} to enrich the dataset.

\begin{table}[t]
\centering
\caption{Statistics of the user-item interaction datasets (after 5-core filtering).}
\label{tab: datasetstat}
\begin{tabular}{lccccc}
\toprule
Dataset & \#Users & \#Items & \#Interactions & Avg. Length & Density \\
\midrule
Beauty & 22,363 & 12,101 & 198,502 & 8.8 & 0.07\% \\
Sports & 25,598 & 18,357 & 296,337 & 8.3 & 0.05\% \\
Toys   & 19,412 & 19,392 & 167,597 & 8.6 & 0.04\% \\
ML-20m & 96,726 & 16,297 & 1,856,746 & 19.2 & 0.11\% \\
\bottomrule
\end{tabular}
\end{table}

\begin{table}[t]
\centering
\caption{Statistics of the extracted facet attributes used for pre-training.}
\label{tab: facet_stat}
\begin{tabular}{l|l|l|c}
\toprule
\textbf{Dataset} & \textbf{Facet Name} & \textbf{Preprocessing / Selection Logic} & \textbf{\# Classes} \\
\midrule
\multirow{3}{*}{Beauty} & Category & Level-2 Category (e.g., \textit{Skin Care}) & 6 \\
 & Brand & Raw Brand Name (e.g., \textit{L'Oreal}) & 2,076 \\
 & Price & Discretized into 10 intervals & 10 \\
\midrule
\multirow{3}{*}{Sports} & Category & Level-2 Category (e.g., \textit{Cycling}) & 34 \\
 & Brand & Raw Brand Name (e.g., \textit{Coleman}) & 2,411 \\
 & Price & Discretized into 10 intervals & 10 \\
\midrule
\multirow{3}{*}{Toys} & Category & Level-2 Category (e.g., \textit{Action Figures}) & 23 \\
 & Brand & Raw Brand Name (e.g., \textit{LEGO}) & 1,312 \\
 & Price & Discretized into 10 intervals & 10 \\
\midrule
\multirow{3}{*}{ML-20m} & Genre & Primary Genre (First in list) & 19 \\
 & Director & Primary Director (First in list) & 3,480 \\
 & Cast & Primary Actor/Actress (First in list) & 3,592 \\
\bottomrule
\end{tabular}
\end{table}

\paragraph{Facet Extraction and Statistics.}
Beyond the raw text, we extract discrete labels for specific facets to drive the supervised contrastive learning (Section~\ref{subsec:supcon}). The statistics and preprocessing logic for these facets are detailed in Table~\ref{tab: facet_stat}.

\begin{itemize}[left=0pt]
    \item \textbf{Amazon Facets (Category, Brand, Price):} 
    The category structure in Amazon is hierarchical (e.g., \textit{"Beauty, Makeup, Nails, Nail Polish"}). To balance granularity, we extract the \textbf{Level-2 Category} (e.g., \textit{Makeup}, \textit{Skin Care}) as the category label. Brands are used directly as distinct classes. For price, we discretize the continuous values into \textbf{10 distinct bins} based on value ranges (e.g., \$0-\$50, \$50-\$100, etc.) to form categorical price classes.
    
    \item \textbf{ML-20m Facets (Genre, Director, Cast):} 
    Movies often possess multiple genres, directors, or cast members. To assign a deterministic label for facet clustering, we select the \textbf{primary} (first-listed) attribute for each field. For example, if a movie's genre list is \textit{["Action", "Adventure", "Sci-Fi"]}, it is assigned the label \textit{Action} for the genre facet. This yields 19 genre classes, 3,480 director classes, and 3,592 cast classes, ensuring robust coverage for contrastive alignment.
\end{itemize}

\subsection{Evaluation Metrics}
We rank the prediction on the whole item set without negative sampling~\cite{fullsort}. Performance is evaluated on a variety of evaluation metrics, including Hit Ratio@$k$ (HR@$k$), and Normalized Discounted Cumulative Gain@$k$ (NDCG@$k$) where $k\in \{5, 10, 20\}$.
Following standard practice in sequential recommendation~\cite{ren2020sequential, SASRec, BERT4Rec, S3-Rec}, we employ a \textit{leave-one-out} evaluation strategy: for each user sequence, the final item serves as the test data, the penultimate item as the validation data, and the remaining items as the training data.

\subsection{Baselines}
We compare our proposed method against a set of baseline models as follows:

\begin{itemize}[left=0pt]
        \item \textbf{SASRec}~\cite{SASRec}: this method is a pioneering work utilizing self-attention to capture dynamic user interests.
        \item \textbf{BERT4Rec}~\cite{BERT4Rec}: this approach adapts the BERT architecture for sequential recommendation using a cloze task.
        \item \textbf{CORE}~\cite{CORE}: it proposes a representation-consistent encoder based on linear combinations of item embeddings to ensure that sequence representations are in the same space with item embeddings.
        \item \textbf{CL4SRec}~\cite{CL4SRec}: this method combines contrastive learning with a Transformer-based model through data augmentation techniques (i.e., item crop, mask, and reorder).
        \item \textbf{ICLRec}~\cite{ICLRec}: this approach improves sequential recommendation by conducting clustering and contrastive learning on user intentions represented by cluster centroids to enhance recommendation.
        \item \textbf{DuoRec}~\cite{DuoRec}: this research investigates the representation degeneration issue in sequential recommendation and offers solutions based on contrastive learning techniques.
        \item \textbf{MSGIFSR}~\cite{MSGIFSR}: it captures multi-level user intents using a Multi-granularity Intent Heterogeneous Session Graph. 
        \item \textbf{Atten-Mixer}~\cite{atten-mixer}: this method leverages concept-view and instance-view readouts for multi-level intent reasoning instead of using the GNN propagation.
        \item \textbf{MiasRec}~\cite{MiasRec}: this approach utilizes multiple item representations in the sequence instead of only using the last item's representation as the sequence representation to capture diverse user intents.
\end{itemize}

\subsection{Settings and Implementation Details}

\subsubsection{Baseline Implementations}
We employ the official source codes for SASRec, ICLRec, MSGIFSR, Atten-Mixer, and MiaSRec. For GRU4Rec and CORE, we leverage the RecBole library\footnote{https://github.com/RUCAIBox/RecBole}~\cite{recbole[1.2.0]}, while BERT4Rec, CL4SRec, and DuoRec are implemented using the SSLRec framework\footnote{https://github.com/HKUDS/SSLRec}~\cite{SSLRec}. All baseline hyperparameters are initialized according to their respective original papers. We explore embedding dimensions of $\{64, 128\}$ and report the results from the configuration that yields the best performance for each model, noting that larger dimensions occasionally led to convergence instability on sparse datasets.

\subsubsection{Text-Enhanced Pre-training}
For the text-enhanced pre-training module, we utilize the pre-trained \textbf{BERT}~\cite{BERT} model (e.g., \texttt{bert-base-uncased}) as the frozen text encoder. To generate robust initial item representations, we apply \textbf{average pooling} over the output sequence tokens, effectively aggregating the semantic information from the item descriptions.

To ensure structural consistency with the downstream FAME architecture, we instantiate the projector with $H=2$ independent heads, corresponding to the two primary facets selected for each dataset (e.g., Category and Brand for Amazon datasets, as detailed in Section~\ref{sec: ablation}). To address the challenge of label sparsity in contrastive learning, we implement a \textbf{Stratified $P \times K$ Sampler}. Specifically, we set $P=4$ and $K=8$, ensuring that every mini-batch contains exactly 4 distinct classes with 8 samples each, thereby guaranteeing valid positive pairs for the loss calculation. The pre-training is conducted for 300 epochs using the alternating optimization strategy described in Section~\ref{sec:method_extension}.

\subsubsection{FAME Optimization}
Our proposed method is implemented in PyTorch. The model is optimized using the Adam optimizer with a learning rate of $0.001$, $\beta_1 = 0.9$, and $\beta_2 = 0.999$. We employ a batch size of 256 for sequential training. For the specific FAME hyperparameters, the number of heads $H$ and the number of experts $N$ are tuned via grid search within the ranges $\{1, 2, 4, 8, 16\}$ and $\{2, 4, 8, 16, 32\}$, respectively. All experiments were conducted on a workstation equipped with a single NVIDIA RTX A5000 GPU.

\begin{table}[htb]
\small
\centering
\caption{Performance comparison of different methods on top-$k$ recommendation}
\label{tab: mainresults}
\begin{tabular}{clccccccccccc}
\toprule
Dataset                 & Metric  & SASRec & BERT4Rec     & CORE   & CL4SRec & ICLRec       & DuoRec       & A-Mixer & MSGIFSR & MiaSRec & FAME+           & Improv. \\
\midrule
\multirow{6}{*}{Beauty} & HR@5    & 0.0508 & 0.0510       & 0.0331 & 0.0623  & {\ul 0.0664} & 0.0504       & 0.0507  & 0.0518  & 0.0524  & \textbf{0.0720} & 8.58\%  \\
                        & HR@10   & 0.0761 & 0.0745       & 0.0664 & 0.0877  & {\ul 0.0918} & 0.0691       & 0.0752  & 0.0771  & 0.0795  & \textbf{0.0993} & 8.17\%  \\
                        & HR@20   & 0.1057 & 0.1075       & 0.1071 & 0.1195  & {\ul 0.1252} & 0.0912       & 0.1033  & 0.1105  & 0.1125  & \textbf{0.1369} & 9.35\%  \\
                        & NDCG@5  & 0.0318 & 0.0343       & 0.0164 & 0.0440  & {\ul 0.0480} & 0.0363       & 0.0350  & 0.0344  & 0.0362  & \textbf{0.0511} & 6.48\%  \\
                        & NDCG@10 & 0.0400 & 0.0419       & 0.0271 & 0.0521  & {\ul 0.0562} & 0.0424       & 0.0421  & 0.0429  & 0.0449  & \textbf{0.0597} & 6.25\%  \\
                        & NDCG@20 & 0.0474 & 0.0502       & 0.0373 & 0.0601  & {\ul 0.0646} & 0.0479       & 0.0504  & 0.0508  & 0.0532  & \textbf{0.0709} & 9.75\%  \\
\hline
\multirow{6}{*}{Sports} & HR@5    & 0.0266 & 0.0252       & 0.0150 & 0.0338  & {\ul 0.0384} & 0.0225       & 0.0217  & 0.0268  & 0.0270  & \textbf{0.0410} & 6.78\%  \\
                        & HR@10   & 0.0412 & 0.0395       & 0.0342 & 0.0498  & {\ul 0.0543} & 0.0327       & 0.0321  & 0.0425  & 0.0435  & \textbf{0.0585} & 7.73\%  \\
                        & HR@20   & 0.0618 & 0.0607       & 0.0609 & 0.0723  & {\ul 0.0753} & 0.0476       & 0.0469  & 0.0634  & 0.0651  & \textbf{0.0831} & 10.34\% \\
                        & NDCG@5  & 0.0158 & 0.0166       & 0.0072 & 0.0235  & {\ul 0.0266} & 0.0161       & 0.0165  & 0.0171  & 0.018   & \textbf{0.0287} & 7.89\%  \\
                        & NDCG@10 & 0.0205 & 0.0212       & 0.0134 & 0.0287  & {\ul 0.0317} & 0.0193       & 0.0188  & 0.0221  & 0.0233  & \textbf{0.0346} & 9.15\%  \\
                        & NDCG@20 & 0.0256 & 0.0265       & 0.0201 & 0.0344  & {\ul 0.0370} & 0.0231       & 0.0235  & 0.0279  & 0.0288  & \textbf{0.0418} & 12.97\% \\
\hline
\multirow{6}{*}{Toys}   & HR@5    & 0.0489 & 0.0464       & 0.0338 & 0.0658  & {\ul 0.0792} & 0.0481       & 0.0565  & 0.0576  & 0.0581  & \textbf{0.0838} & 5.78\%  \\
                        & HR@10   & 0.0676 & 0.0677       & 0.0699 & 0.0912  & {\ul 0.1043} & 0.0666       & 0.0819  & 0.0831  & 0.0828  & \textbf{0.1098} & 5.27\%  \\
                        & HR@20   & 0.0908 & 0.0968       & 0.1114 & 0.1209  & {\ul 0.1382} & 0.0879       & 0.1099  & 0.1150  & 0.1143  & \textbf{0.1437} & 3.98\%  \\
                        & NDCG@5  & 0.0329 & 0.0322       & 0.0158 & 0.0471  & {\ul 0.0579} & 0.0356       & 0.0403  & 0.0407  & 0.0408  & \textbf{0.0618} & 6.74\%  \\
                        & NDCG@10 & 0.0389 & 0.0391       & 0.0274 & 0.0552  & {\ul 0.0660} & 0.0415       & 0.0481  & 0.0492  & 0.0488  & \textbf{0.0712} & 7.88\%  \\
                        & NDCG@20 & 0.0448 & 0.0464       & 0.0378 & 0.0627  & {\ul 0.0745} & 0.0469       & 0.0574  & 0.0577  & 0.0567  & \textbf{0.0796} & 6.85\%  \\
\hline
\multirow{6}{*}{ML-20m} & HR@5    & 0.1305 & 0.1446       & 0.0655 & 0.1205  & {\ul 0.1380} & 0.1458       & 0.1325  & 0.1303  & 0.1367  & \textbf{0.1588} & 9.82\%  \\
                        & HR@10   & 0.2016 & 0.2172       & 0.1312 & 0.1853  & 0.2070       & {\ul 0.2164} & 0.2022  & 0.2013  & 0.2071  & \textbf{0.2373} & 9.66\%  \\
                        & HR@20   & 0.2996 & {\ul 0.3132} & 0.2251 & 0.2760  & 0.2997       & 0.3108       & 0.2994  & 0.2978  & 0.3021  & \textbf{0.3397} & 8.46\%  \\
                        & NDCG@5  & 0.0858 & {\ul 0.0964} & 0.0347 & 0.0804  & 0.0927       & 0.0986       & 0.0899  & 0.0844  & 0.0918  & \textbf{0.1068} & 8.32\%  \\
                        & NDCG@10 & 0.1086 & 0.1197       & 0.0558 & 0.1012  & 0.1149       & {\ul 0.1212} & 0.1121  & 0.1119  & 0.1144  & \textbf{0.1363} & 12.46\% \\
                        & NDCG@20 & 0.1333 & 0.1438       & 0.0794 & 0.124   & 0.1382       & {\ul 0.1450} & 0.1334  & 0.1357  & 0.1383  & \textbf{0.1608} & 11.85\% \\
\bottomrule
\end{tabular}
\end{table}
\subsection{Overall Performance}

Table~\ref{tab: mainresults} presents the comprehensive performance comparison of our proposed framework against various state-of-the-art baselines. Note that we omit GRU4Rec due to space constraints, as transformer-based models generally demonstrate superior capability in sequential modeling. Our experimental findings yield several critical observations:

\begin{itemize}[left=0pt]
    \item \textbf{Performance of General Sequential Models:} 
    Standard self-attention architectures (e.g., SASRec, BERT4Rec) achieve competitive performance by capturing long-term dependencies. However, their reliance on ID-based embeddings limits their ability to capture fine-grained semantic relationships, often resulting in suboptimal performance compared to models that explicitly model latent intents or item features.

    \item \textbf{Efficacy of Intent Disentanglement:} 
    Models designed to capture user intents (e.g., ICLRec, DuoRec) or multi-interest representations (e.g., MiaSRec) consistently outperform single-vector sequential models. This improvement validates the necessity of moving beyond superficial item transitions. Specifically, contrastive learning approaches (e.g., ICLRec, CL4SRec) effectively mitigate noise, while multi-interest frameworks (e.g., MiaSRec) better handle diverse user behaviors. Nevertheless, these methods primarily focus on interaction patterns and often neglect the inherent multi-faceted semantics of the items themselves.

    \item \textbf{Superiority of FAME+:} 
    Our proposed \textbf{FAME+} (FAME with Text-Enhanced Facet-Aware Pre-training) achieves the best performance across all datasets and metrics, significantly outperforming the strongest baselines with improvements ranging from \textbf{3.98\% to 12.97\%}. This empirical superiority is driven by two key factors:
    \begin{enumerate}
        \item \textbf{Granular Preference Modeling:} The MoE architecture in the FAME backbone successfully disentangles complex user preferences within specific facets, allowing for more precise next-item prediction.
        \item \textbf{Semantic Alignment via Text Enhancement:} The text-enhanced pre-training module is particularly effective. By explicitly extracting and disentangling facet features (e.g., Genre, Director) via supervised contrastive learning, FAME+ overcomes the cold-start limitations of ID-based embeddings. This is especially evident in the \textit{Sports} and \textit{ML-20m} datasets, where the richer semantic initialization leads to substantial gains in NDCG (e.g., \textbf{12.97\%} improvement on Sports NDCG@20), proving that explicitly modeling item facets yields a deeper understanding of user intent.
    \end{enumerate}
\end{itemize}

\subsection{Parameters Study}
This subsection presents the ablation study to evaluate the contributions of our proposed components and conduct corresponding hyperparameter tuning. We begin by examining FAME$_{w/o\; MoE}$, which excludes the MoE module, to assess the impact of the facet-aware multi-head prediction mechanism (introduced in Section.~\ref{MultiHeadPLayer}) and determine the optimal number of attention heads in Section.~\ref{headexp}. Subsequently, using the optimized head configuration, we evaluate the complete FAME model to validate the effectiveness of the MoE module and identify the optimal number of experts in Section.~\ref{expertexp}.
\begin{figure*}[t]
\centering
     \subfloat[Beauty]{\includegraphics[width=0.25\linewidth]{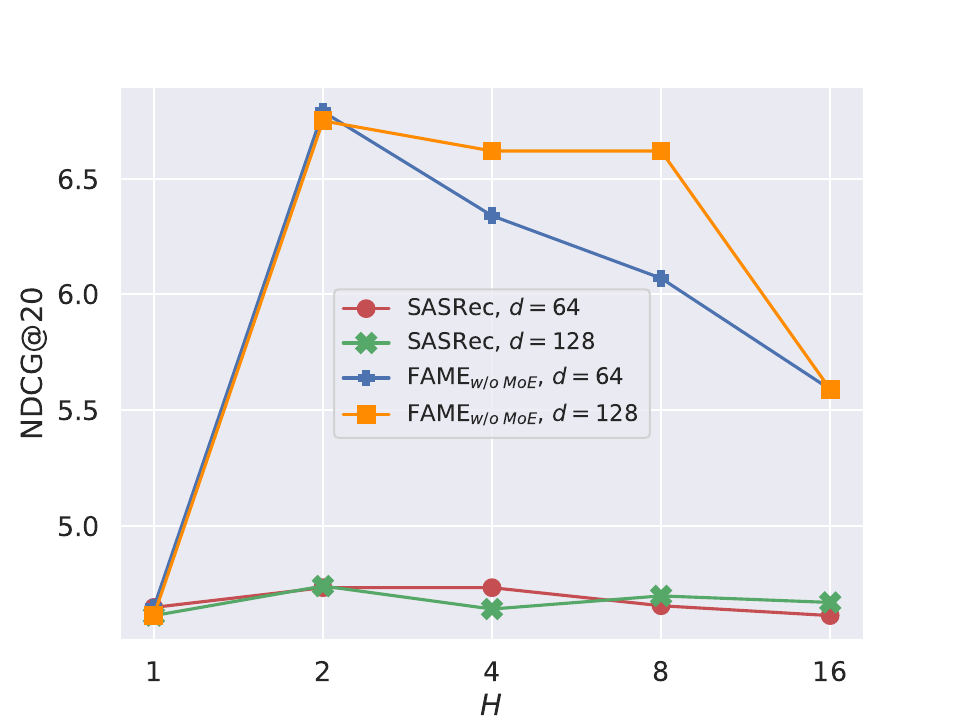}\label{fig: beautyNDCG}}
    \subfloat[Sports]{\includegraphics[width=0.25\linewidth]{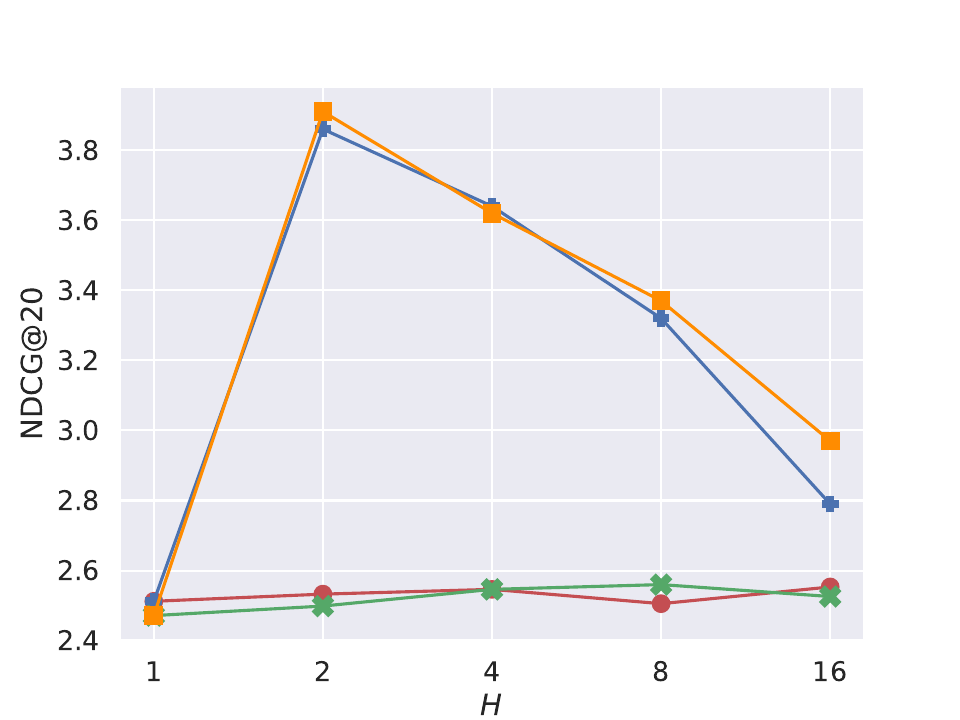}\label{fig: sportsNDCG}}
	\subfloat[Toys]{\includegraphics[width=0.25\linewidth]{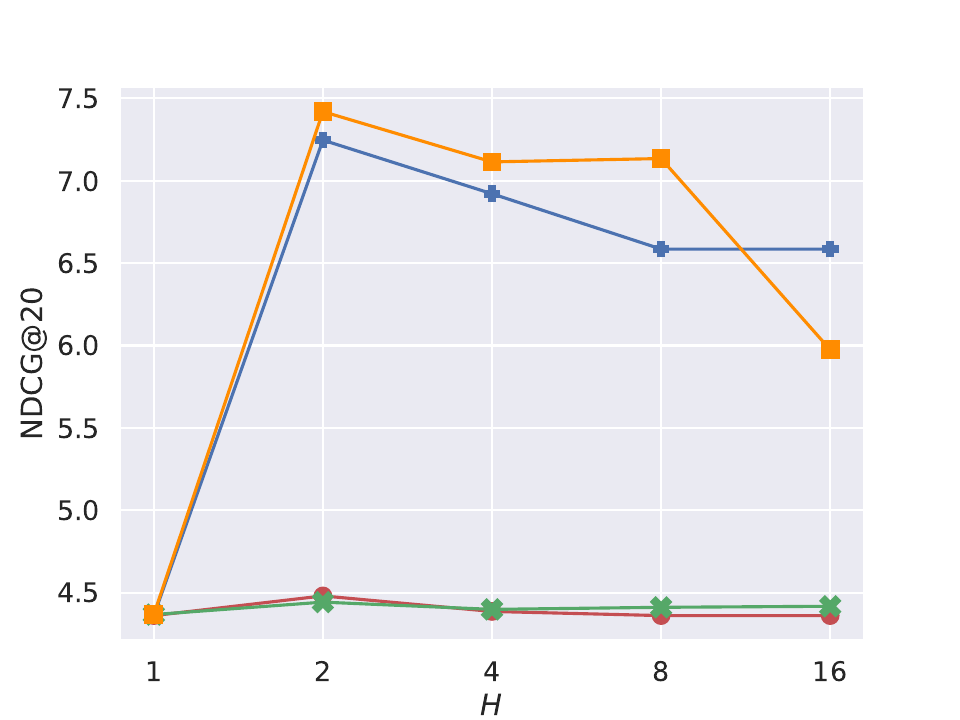}\label{fig: toysNDCG}}
	\subfloat[ML-20m]{\includegraphics[width=0.25\linewidth]{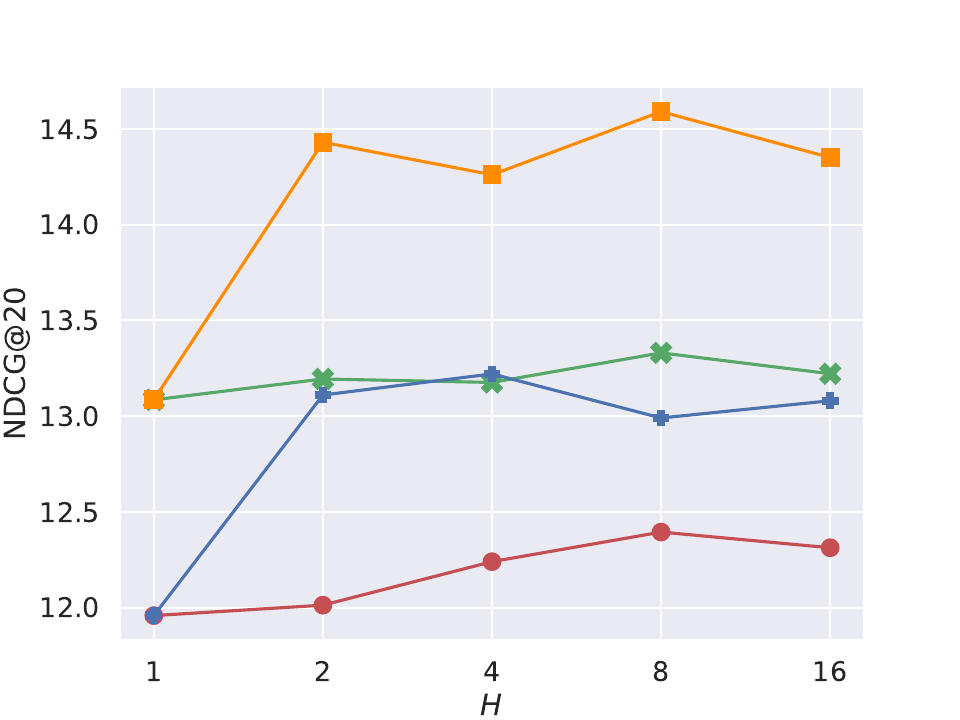}\label{fig: ml20mNDCG}}\\
    \subfloat[Beauty]{\includegraphics[width=0.25\linewidth]{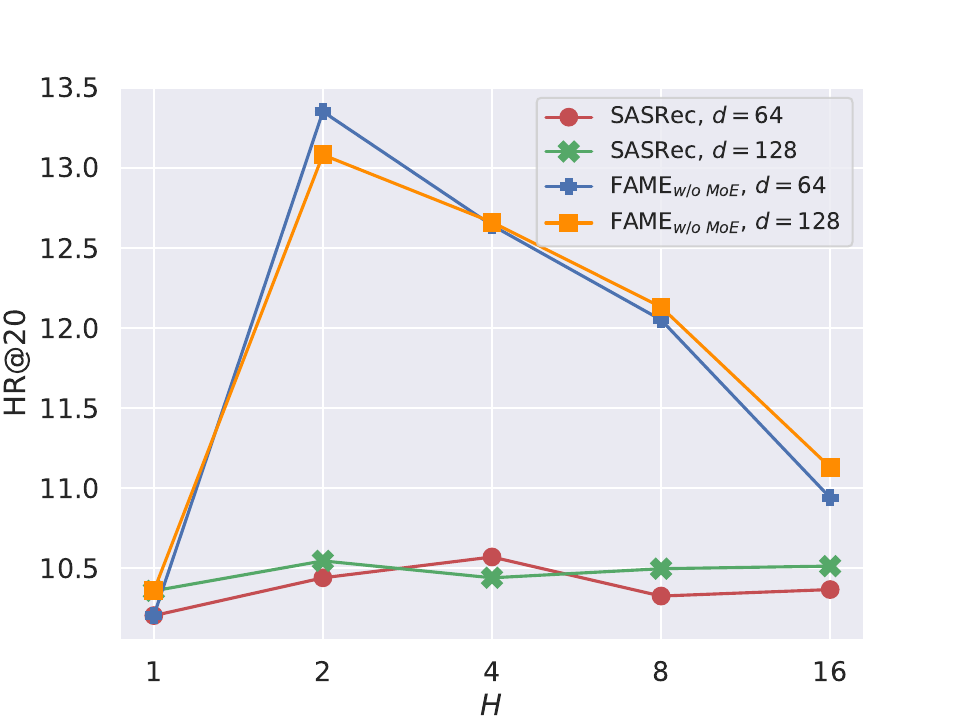}\label{fig: beautyHR}}
    \subfloat[Sports]{\includegraphics[width=0.25\linewidth]{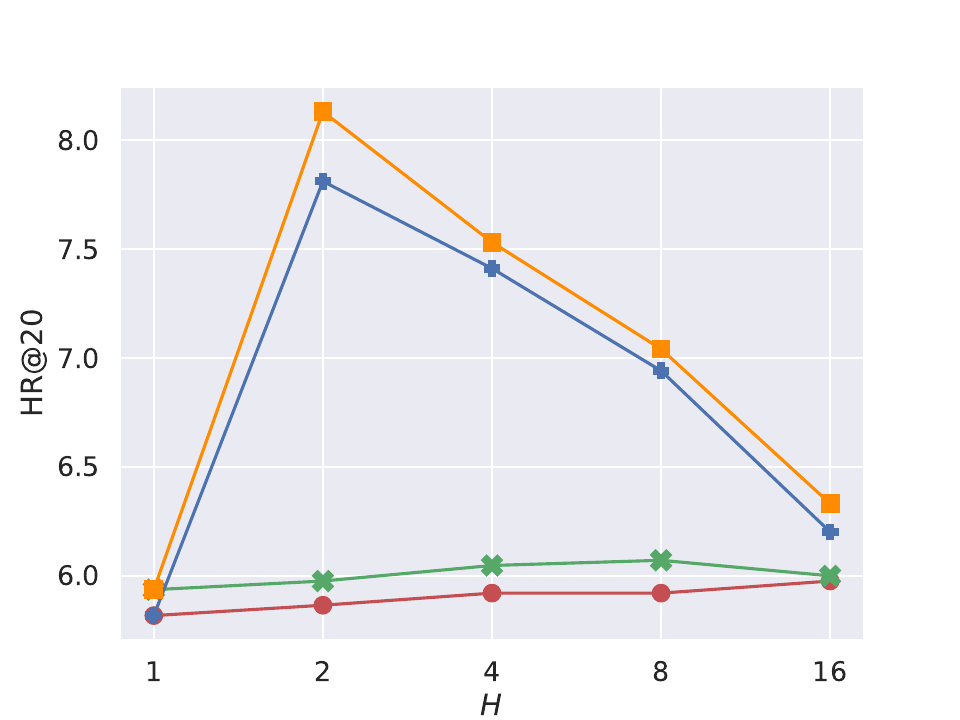}\label{fig: sportsHR}}
	\subfloat[Toys]{\includegraphics[width=0.25\linewidth]{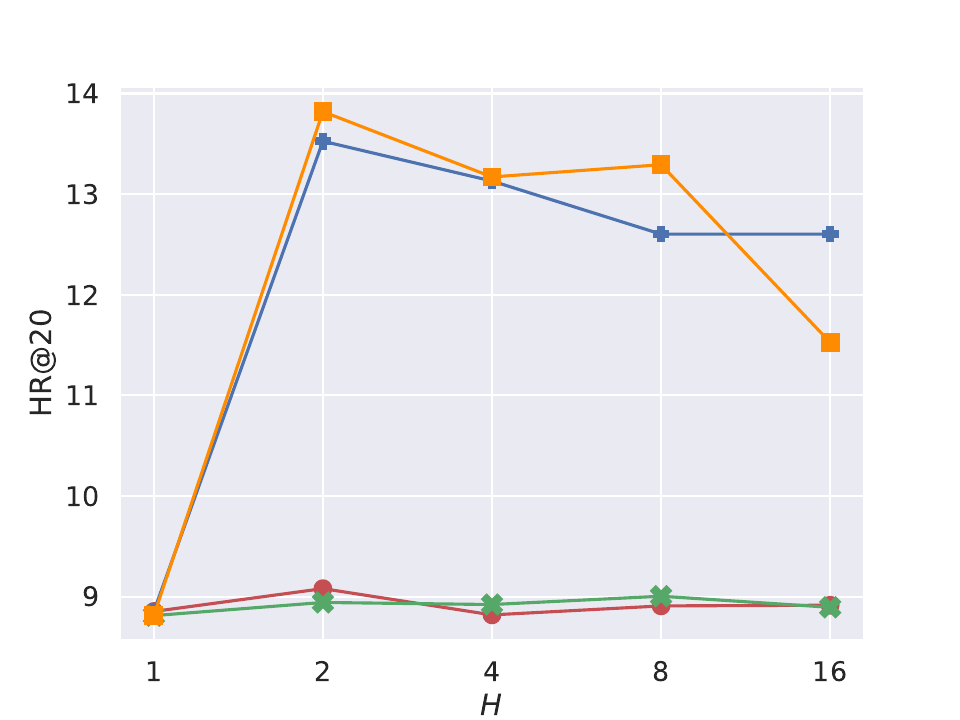}\label{fig: toysHR}}
	\subfloat[ML-20m]{\includegraphics[width=0.25\linewidth]{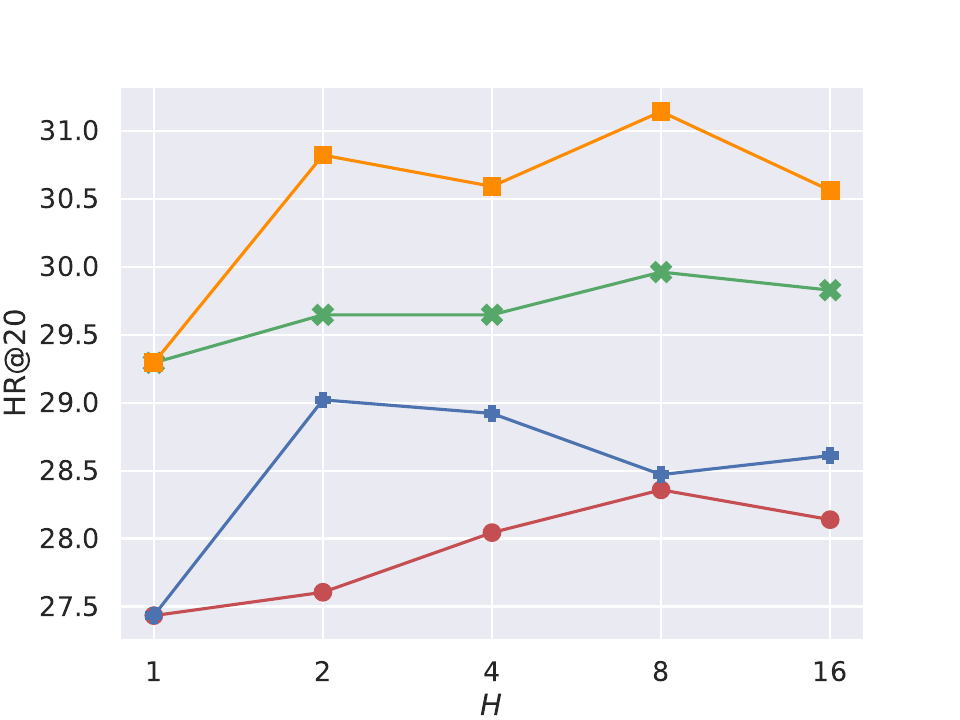}\label{fig: ml20mHR}}
\caption{The performances comparison varying the number of heads in each dataset. The metric in (a)-(d) is NDCG@20, and the metric in (e)-(h) is HR@$20$.}
\label{fig: head study}
\end{figure*}

\begin{figure*}[t]
\centering
     \subfloat[Beauty]{\includegraphics[width=0.25\linewidth]{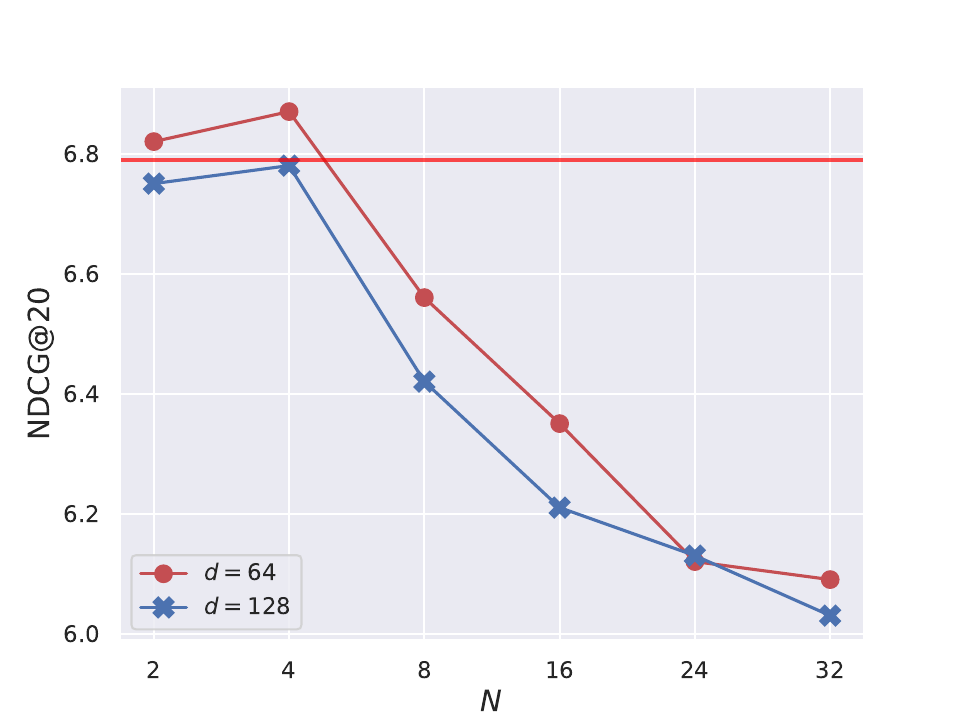}\label{fig: beautyNDCGMoE}}
    \subfloat[Sports]{\includegraphics[width=0.25\linewidth]{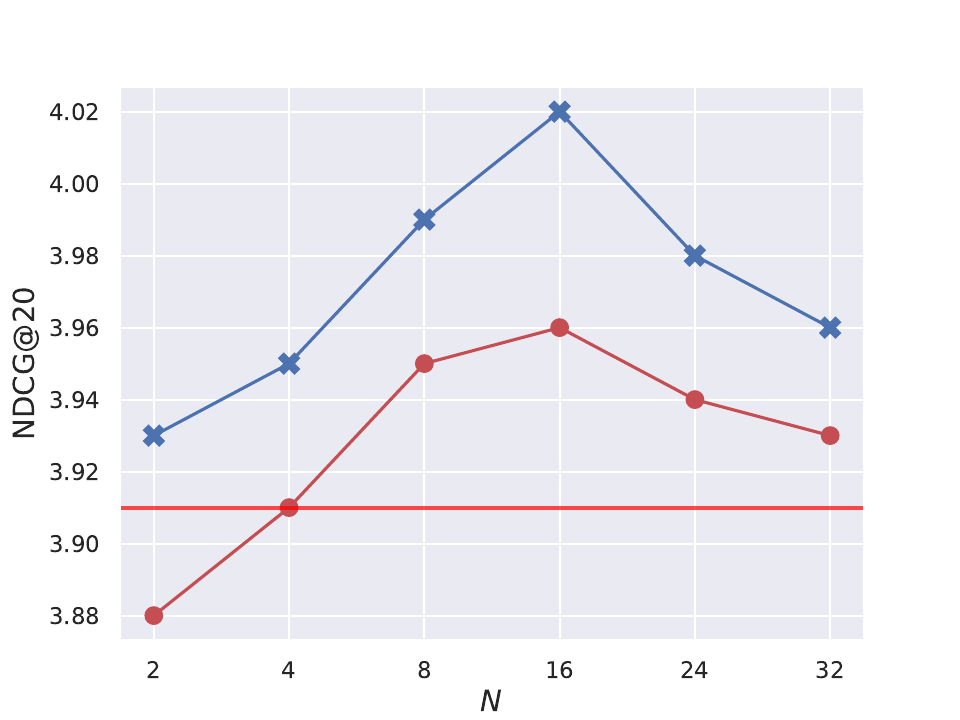}\label{fig: sportsNDCGMoE}}
	\subfloat[Toys]{\includegraphics[width=0.25\linewidth]{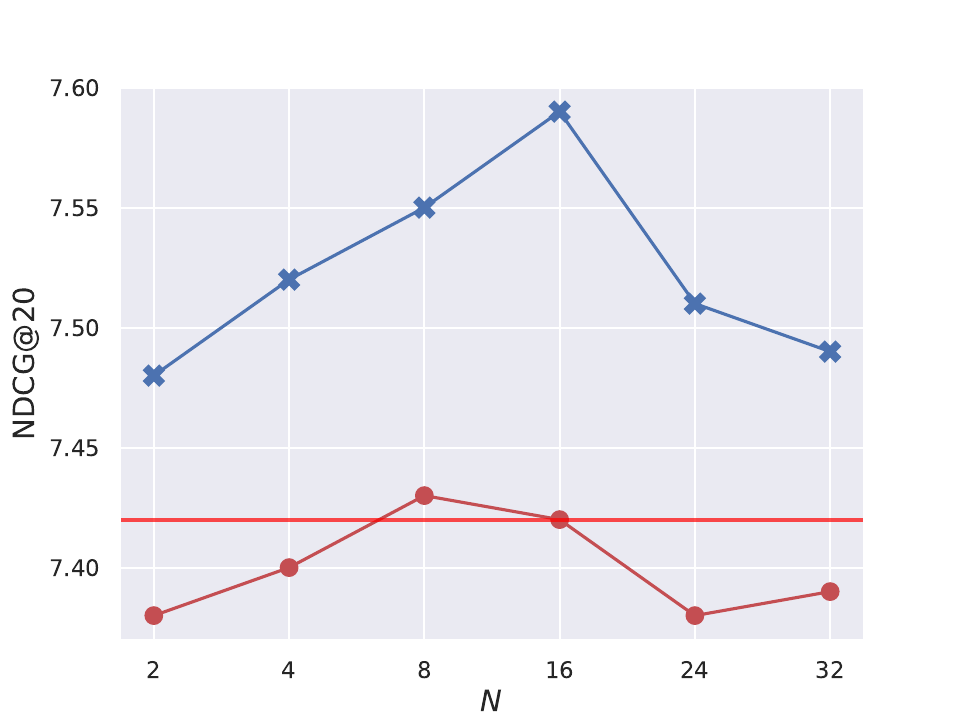}\label{fig: toysNDCGMoE}}
	\subfloat[ML-20m]{\includegraphics[width=0.25\linewidth]{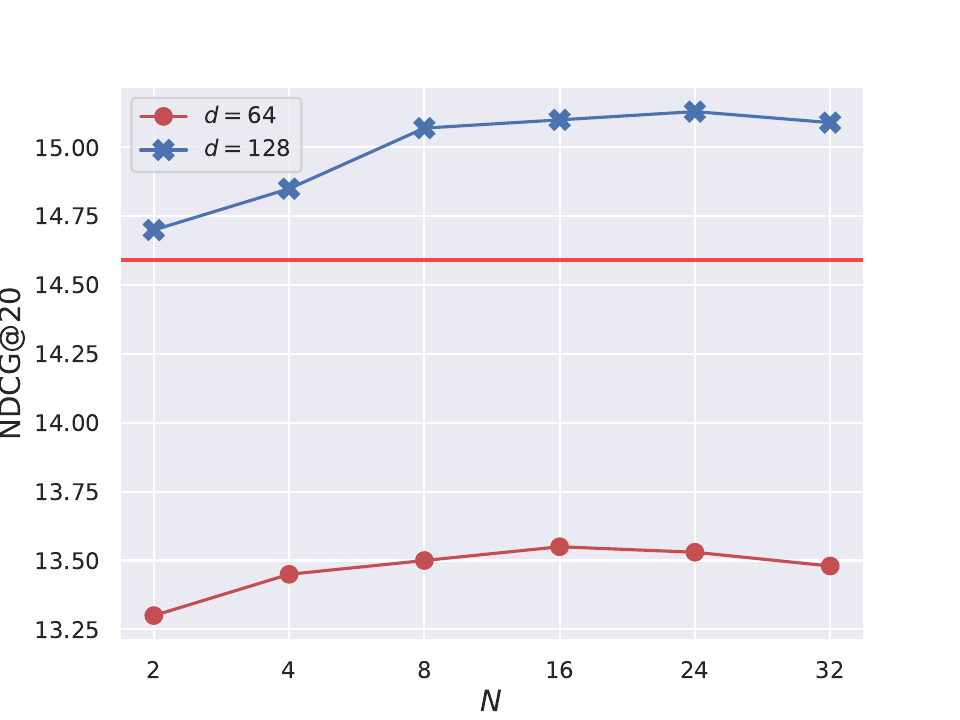}\label{fig: ml20mNDCGMoE}}\\
    \subfloat[Beauty]{\includegraphics[width=0.25\linewidth]{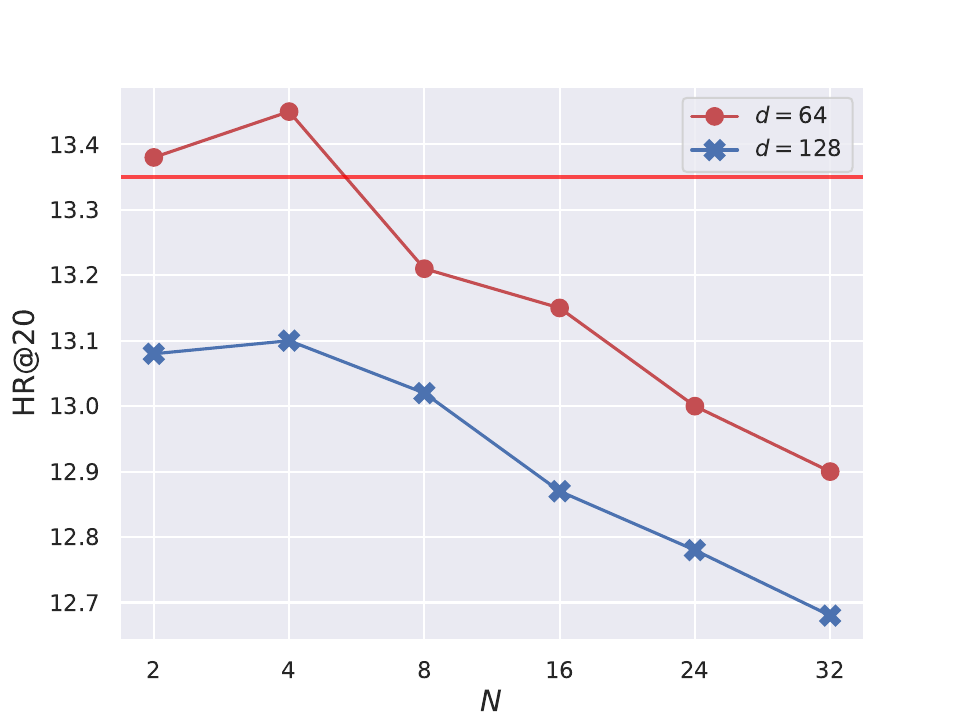}\label{fig: beautyHRMoE}}
    \subfloat[Sports]{\includegraphics[width=0.25\linewidth]{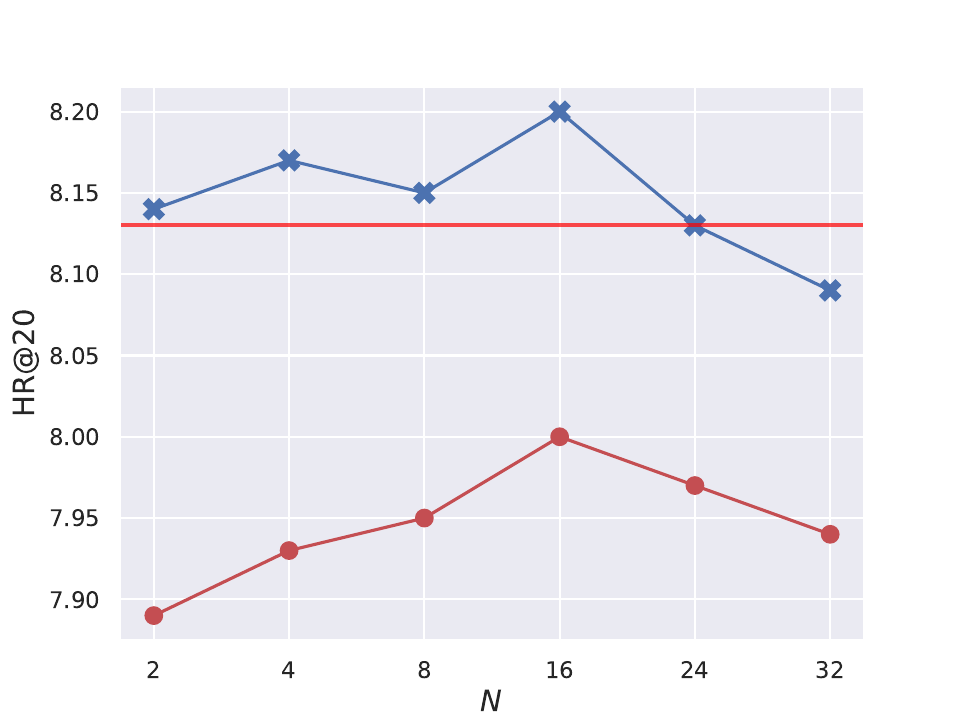}\label{fig: sportsHRMoE}}
	\subfloat[Toys]{\includegraphics[width=0.25\linewidth]{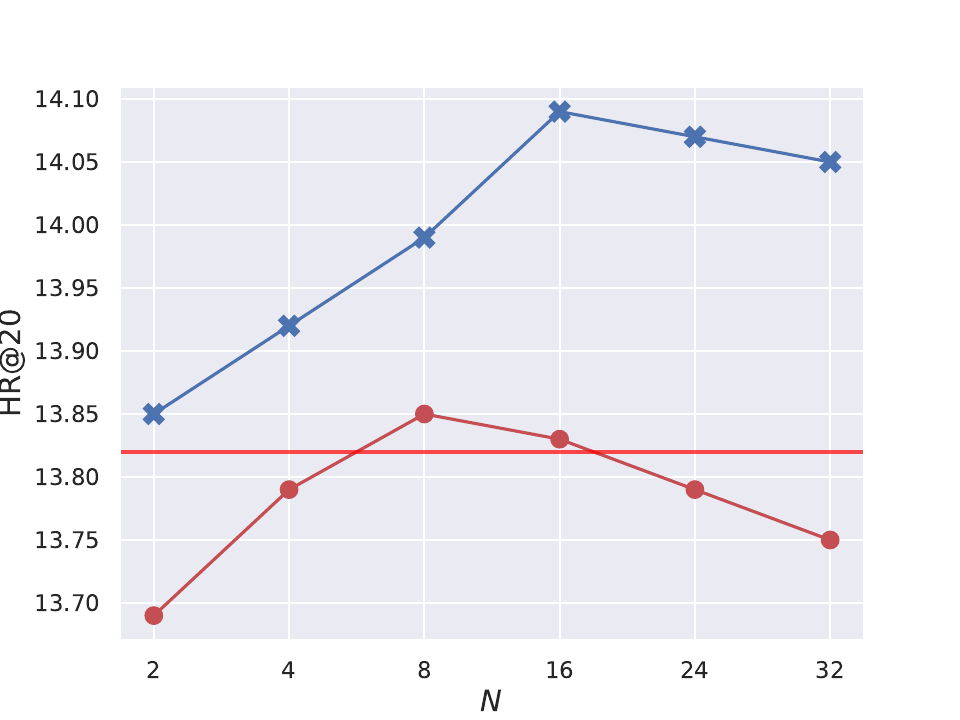}\label{fig: toysHRMoE}}
	\subfloat[ML-20m]{\includegraphics[width=0.25\linewidth]{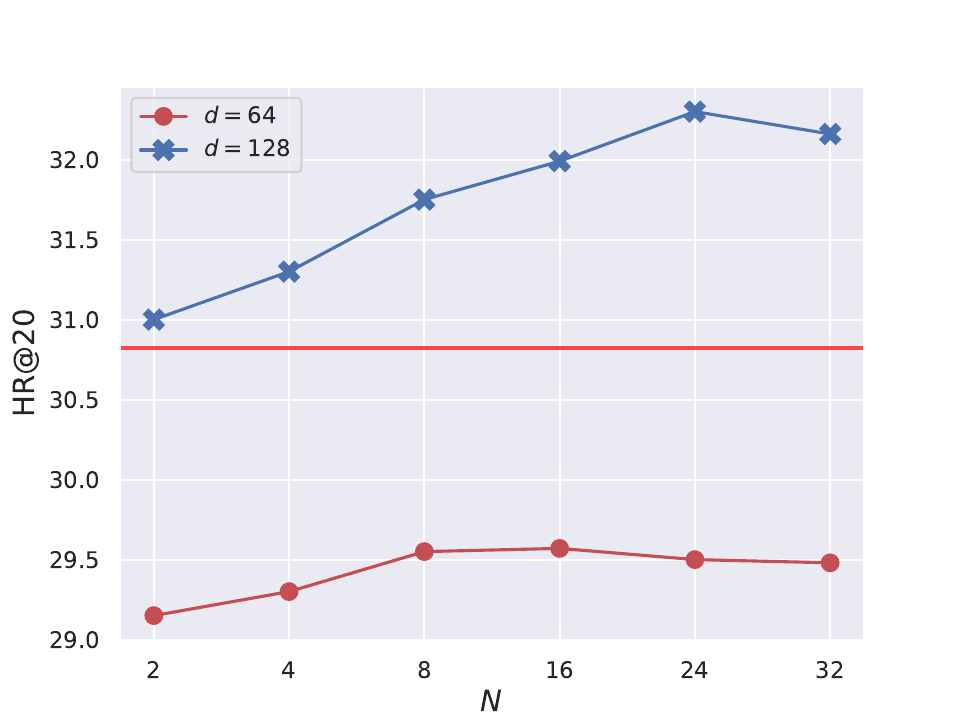}\label{fig: ml20mHRMoE}}
\caption{The performances comparison varying the number of experts in each dataset. The metric in (a)-(d) is NDCG@20, and the metric in (e)-(h) is HR@20. The red horizontal line in each subfigure indicates the peak performance (NDCG@20 or HR@20) achieved by FAME$_{w/o\; MoE}$ within that dataset, as shown in Figure~\ref{fig: head study}.}
\label{fig: expert study}
\end{figure*}

\subsubsection{Impact of the number of heads}
\label{headexp}

Figure~\ref{fig: head study} illustrates the performance variation with different numbers of heads ($H$), treated as a hyperparameter. We compare the original SASRec model with FAME$_{w/o\; MoE}$ to isolate the impact of our multi-head prediction mechanism. 
We experiment with $H$ values of $\{1,2,4,8,16 \}$. When $H=1$, our FAME$_{w/o\; MoE}$ is reduced to the original SASRec model with single head. 
As noted in~\cite{transformer}, computational costs remain constant when varying the number of heads ($H$) while maintaining a fixed embedding dimension ($d$).
\\\textbf{Benefits of multi-head attention:} Both SASRec and FAME$_{w/o\; MoE}$ exhibit performance improvements with multiple heads, however, excessive heads can lead to diminishing returns, aligning with findings in Transformer~\cite{transformer} and SASRec~\cite{SASRec}.
\\\textbf{Superiority of facet-aware architecture:} FAME$_{w/o\; MoE}$ consistently outperforms SASRec, demonstrating the effectiveness of our facet-aware approach.
\\\textbf{Dataset-specific optimal head count:} The optimal number of heads varies across datasets. Beauty, Sports, and Toys benefit from fewer heads, suggesting simpler item facets, while ML-20m requires more heads to capture complex item characteristics.

\subsubsection{Impact of the number of experts}
\label{expertexp}
Figure~\ref{fig: expert study} illustrates the influence of the number of experts ($N$) within each attention head on model performance. We set $H$ to the optimal value determined for FAME$_{w/o\; MoE}$ and compare its performance (red horizontal line in each subfigure) to {\cheng that of} FAME {\cheng by} varying $N$ in $\{2,4,8,16,32 \}$. 
FAME simplifies to FAME$_{w/o\; MoE}$ when $N$ is set to 1.

FAME outperforms FAME$_{w/o\; MoE}$ across all datasets. This improvement can be attributed to the effectiveness of the MoE component, as evidenced by the existence of an optimal $N$ value in each subfigure that surpasses the performance of FAME$_{w/o\; MoE}$.
While the Beauty dataset exhibits diminishing returns for $N$ greater than 4, suggesting simpler user preferences, the other datasets benefit from a larger number of experts. In particular, ML-20m show performance gains with increasing $N$, indicating the presence of more complex and diverse user preferences. However, excessive experts ($N=32$) might lead to overfitting in the Sports and Toys dataset.

\subsection{Ablation Study}
\label{sec: ablation}

To investigate the individual contributions of the text-enhanced initialization and the explicit facet-aware training objectives, we conduct an ablation study comparing the full \textbf{FAME+} model against several variants. The performance comparison is reported in Table~\ref{tab: ablation}.

\begin{table}[htb]
\centering
\caption{Ablation study on text enhancement and facet selection (HR@20 / NDCG@20).}
\label{tab: ablation}
\begin{tabular}{clccccc}
\toprule
Dataset & Metric & FAME & FAME+$_{raw}$ & FAME+$_{1,2}$ & FAME+$_{1,3}$ & FAME+$_{2,3}$ \\
\midrule
\multirow{2}{*}{Beauty} & HR & 0.1345 & 0.1358 & \textbf{0.1369} & {\ul 0.1361} & 0.1358 \\
 & NDCG & 0.0687 & 0.0701 & \textbf{0.0709} & {\ul 0.0705} & 0.0702 \\
\hline
\multirow{2}{*}{Sports} & HR & 0.0820 & 0.0826 & \textbf{0.0831} & {\ul 0.0829} & 0.0827 \\
 & NDCG & 0.0402 & 0.0411 & \textbf{0.0418} & {\ul 0.0415} & 0.0413 \\
\hline
\multirow{2}{*}{Toys} & HR & 0.1409 & 0.1424 & \textbf{0.1437} & {\ul 0.1431} & 0.1426 \\
 & NDCG & 0.0759 & 0.0783 & \textbf{0.0796} & {\ul 0.0789} & 0.0782 \\
\hline
\multirow{2}{*}{ML-20m} & HR & 0.3230 & 0.3391 & \textbf{0.3397} & {\ul 0.3395} & 0.3228 \\
 & NDCG & 0.1513 & 0.1602 & \textbf{0.1608} & {\ul 0.1606} & 0.1514 \\
\bottomrule
\end{tabular}
\end{table}

\noindent \textbf{Experimental Variants.} We define the following variants for comparison:
\begin{itemize}[left=0pt]
    \item \textbf{FAME:} The original backbone framework initialized with random ID-embeddings, without text enhancement~\cite{FAME}.
    \item \textbf{FAME+$_{raw}$:} Utilizing the pre-trained text encoder (e.g., BERT) to obtain item embeddings, projected via a simple MLP to the recommender dimension. This variant incorporates semantic information but lacks the facet-aware contrastive pre-training.
    \item \textbf{FAME+$_{i,j}$:} The complete model pre-trained with selected facet combinations $(i, j) \in \{1, 2, 3\}$.
    \begin{itemize}
        \item \textit{Amazon Datasets:} Facet 1=Category, Facet 2=Brand, Facet 3=Price.
        \item \textit{ML-20m:} Facet 1=Genre, Facet 2=Director, Facet 3=Cast.
    \end{itemize}
\end{itemize}

\subsubsection{Impact of Textual Initialization}
As shown in Table~\ref{tab: ablation}, \textbf{FAME+$_{raw}$} consistently outperforms the random-initialization baseline (\textbf{FAME}) across all datasets. For instance, on the \textit{ML-20m} dataset, the raw text injection improves NDCG@20 from 0.1513 to 0.1602. This confirms that semantic metadata provides critical side information that helps mitigate the cold-start problem inherent in pure ID-based learning, offering a more robust starting point for sequential modeling.

\begin{figure*}[t]
\centering
    \subfloat[Toys Category (Raw Text)]{\includegraphics[width=0.48\linewidth]{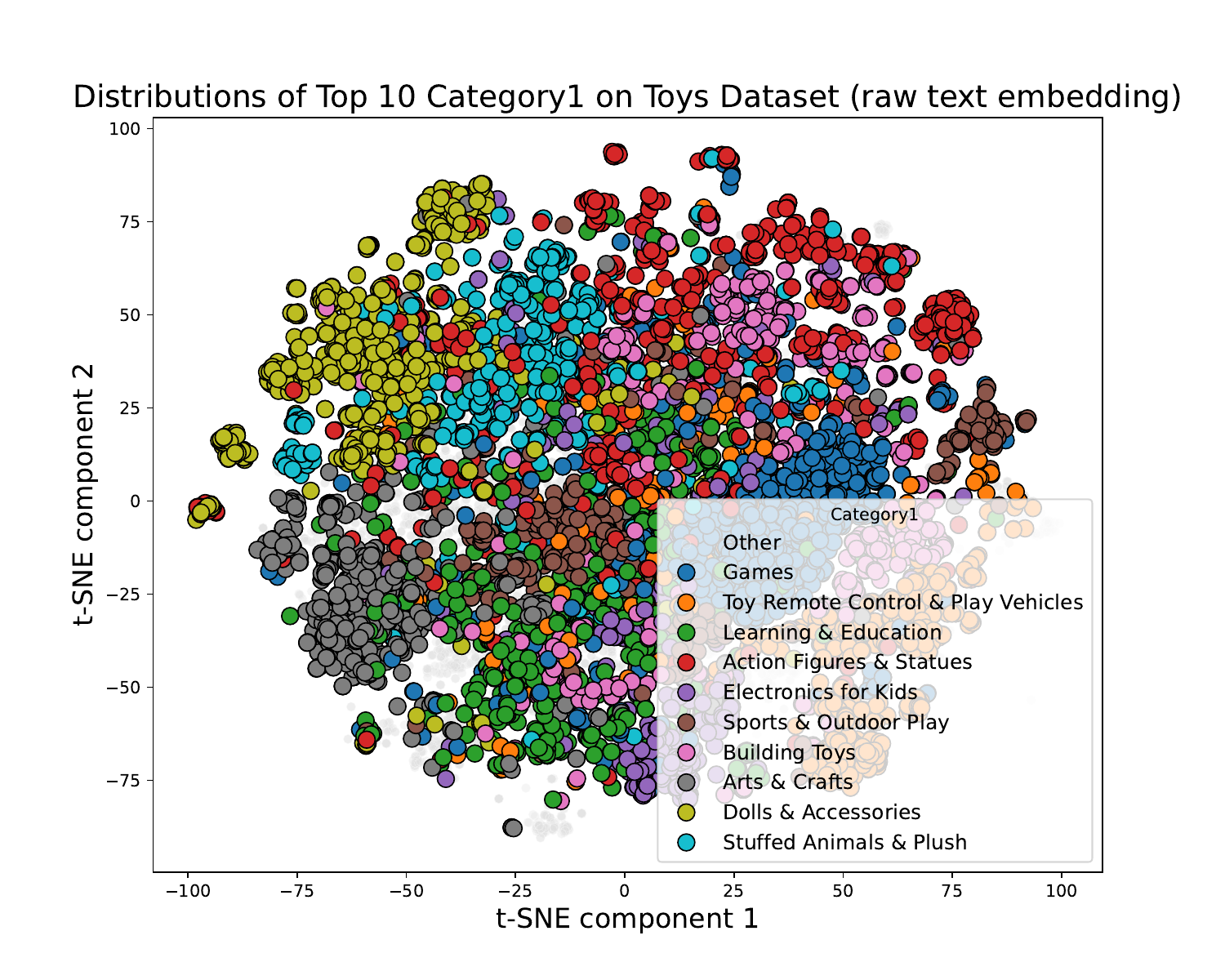}} \hfill
    \subfloat[Toys Category (Facet Subspace)]{\includegraphics[width=0.48\linewidth]{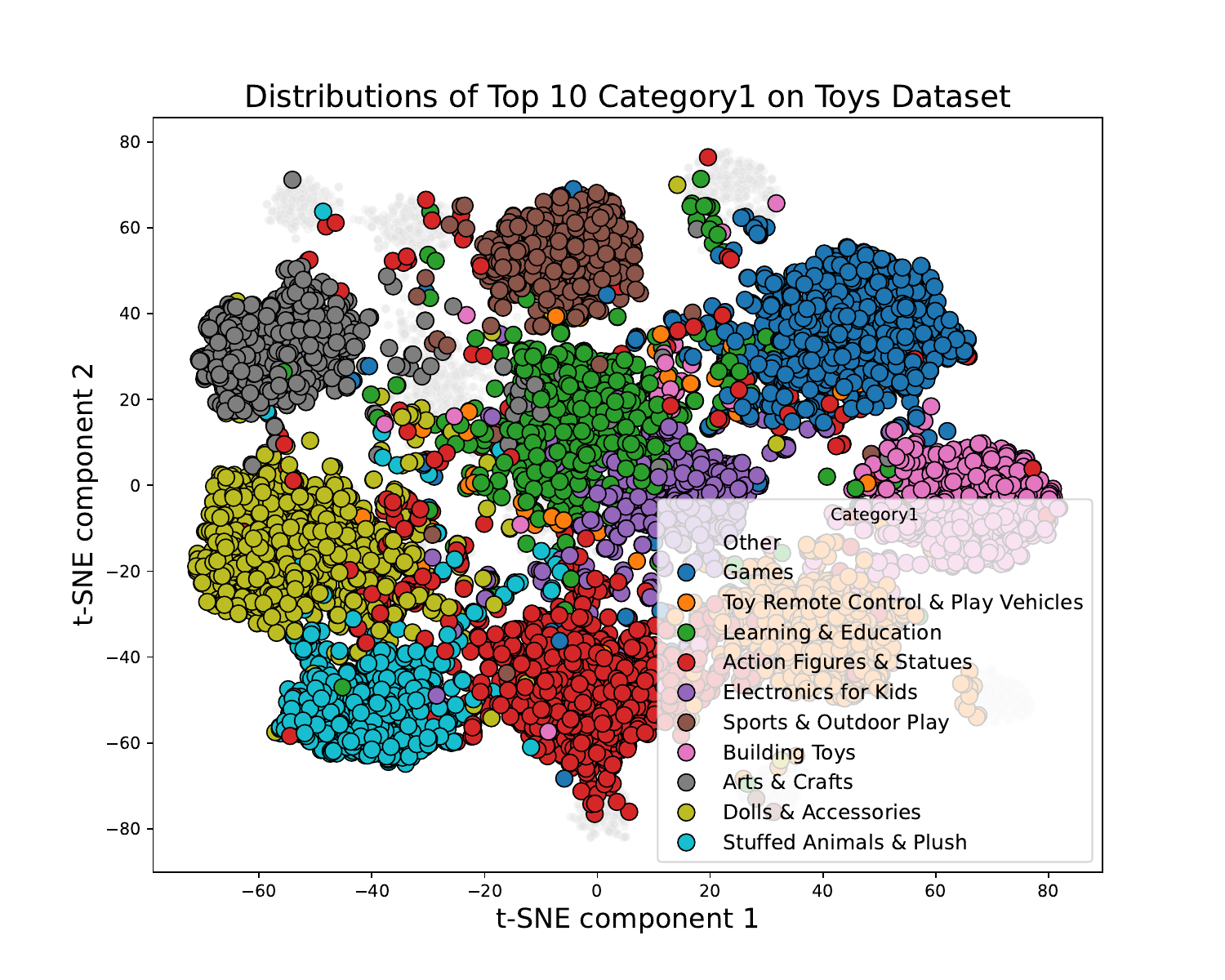}} \\
    \subfloat[ML-20m Genre (Raw Text)]{\includegraphics[width=0.48\linewidth]{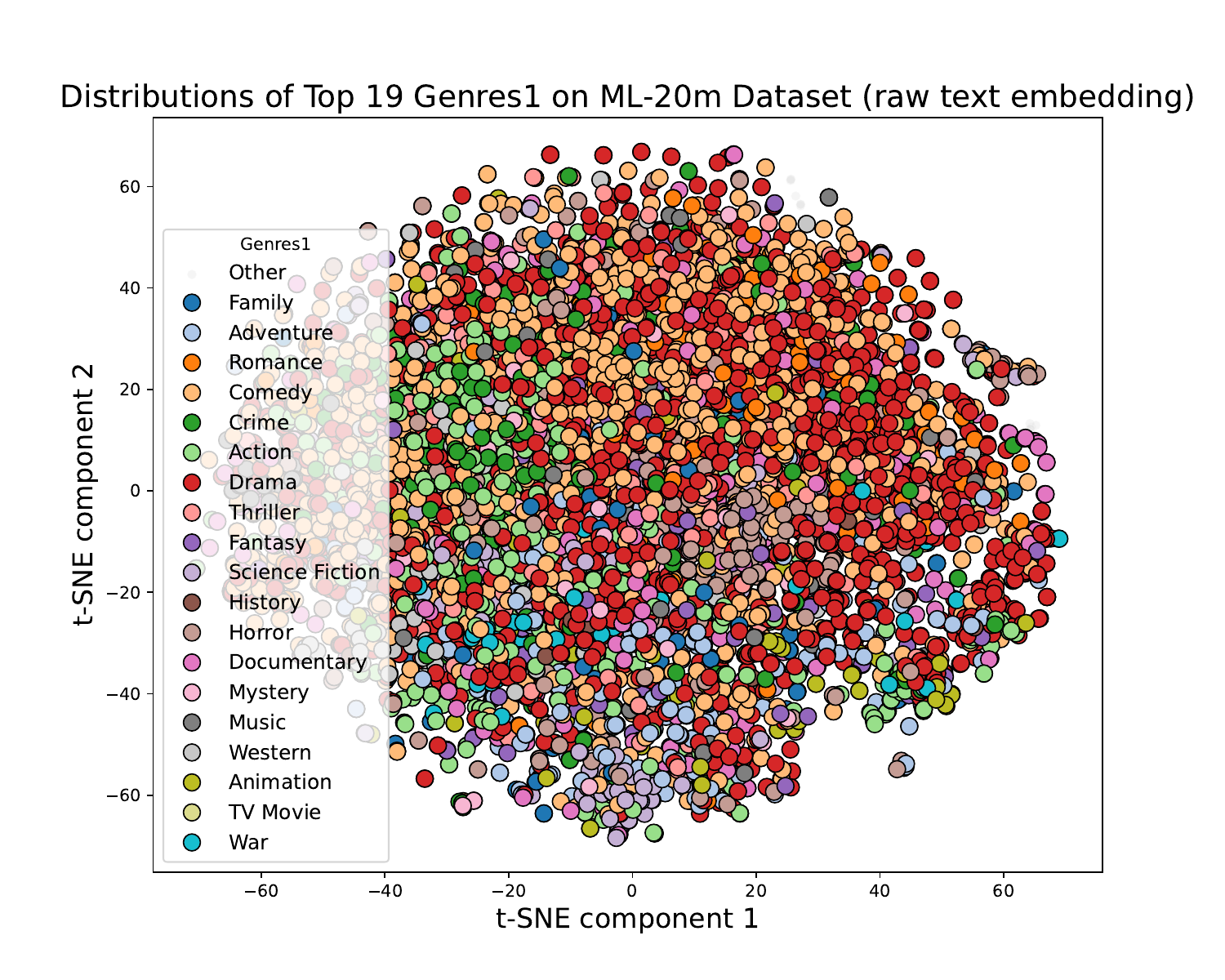}} \hfill
    \subfloat[ML-20m Genre (Facet Subspace)]{\includegraphics[width=0.48\linewidth]{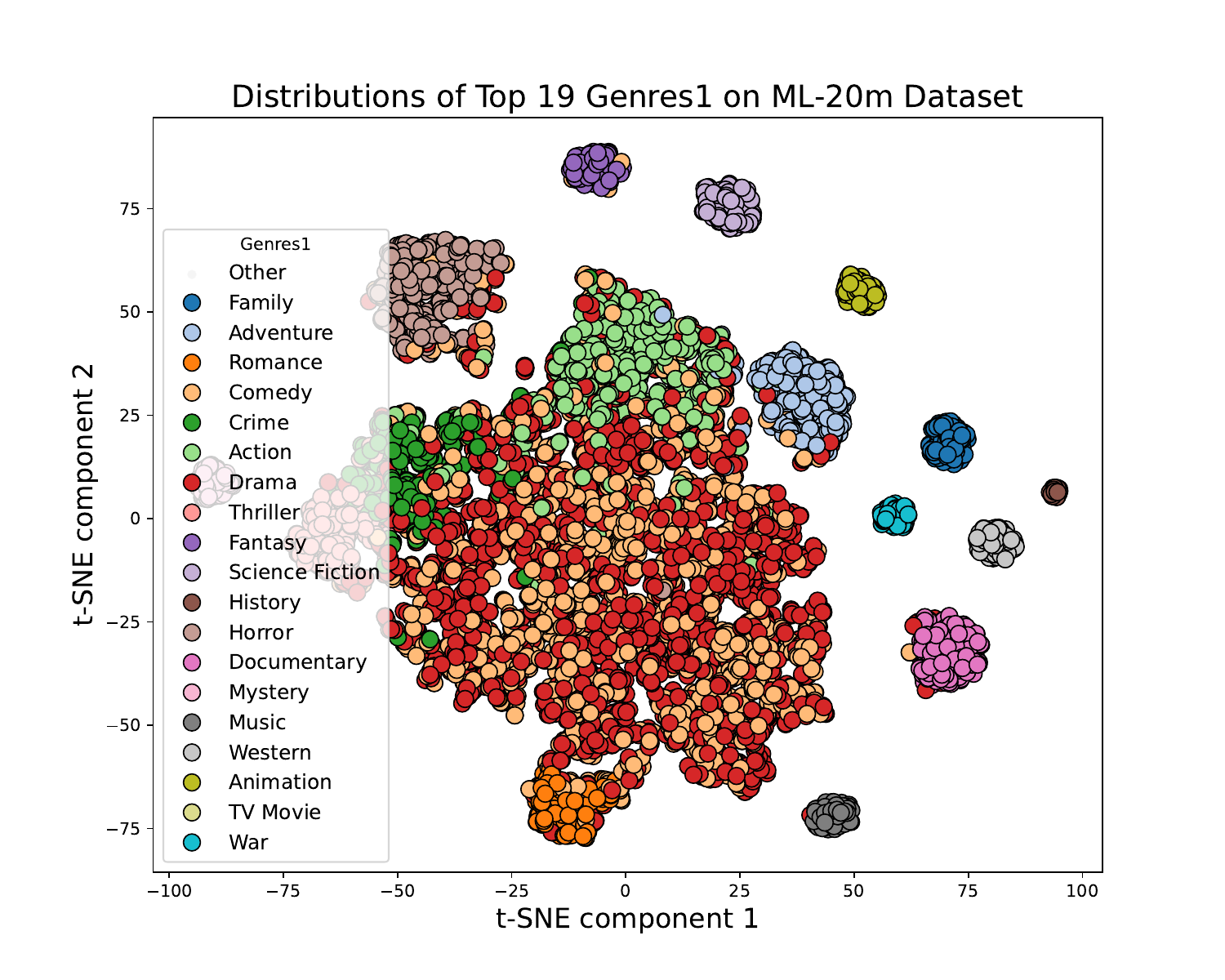}}
\caption{Visualization of item embedding distributions using t-SNE. (a) and (c) show that raw textual embeddings result in entangled distributions. (b) and (d) demonstrate that our facet-aware pre-training successfully disentangles the latent space into distinct semantic clusters (e.g., by Category or Genre).}
\label{fig: visualisation}
\end{figure*}
\subsubsection{Impact of Facet-Aware Disentanglement}
While raw text improves performance, it is insufficient on its own. \textbf{FAME+$_{1,2}$} (our proposed method) consistently achieves the highest performance, surpassing \textbf{FAME+$_{raw}$} across all metrics. This demonstrates that simply ingesting text is not enough; the embeddings must be explicitly disentangled to align with the multi-facet structure of the downstream recommender.

We visualize this effect in Figure~\ref{fig: visualisation}, which plots the t-SNE distributions of item embeddings before and after our pre-training.
\begin{itemize}
    \item \textbf{Before (Raw Text):} As seen in Figures~\ref{fig: visualisation}(a) and (c), the raw BERT embeddings exhibit a centralized, overlapping distribution where distinct categories (e.g., Genres in ML-20m) are inextricably mixed.
    \item \textbf{After (Facet-Aware):} In Figures~\ref{fig: visualisation}(b) and (d), clear, separable clusters emerge corresponding to the ground-truth labels (Category/Genre). This visual evidence confirms that our supervised contrastive learning objective successfully pulls semantically similar items together, creating a structured embedding space that facilitates more accurate retrieval.
\end{itemize}

\subsubsection{Analysis of Facet Importance}
Comparing the different facet combinations reveals that \textbf{Facet 1 (Category/Genre)} is the most impactful driver of recommendation quality. FAME+$_{1,2}$ (Category+Brand/Genre+Director) generally outperforms variants that exclude Facet 1 (e.g., FAME+$_{2,3}$).
\begin{itemize}
    \item \textbf{High-Impact Facets:} Category and Genre represent dense, fundamental user intents, making them crucial for prediction.
    \item \textbf{Sparse/Noisy Facets:} Conversely, attributes like Brand (Amazon) or Director (ML-20m) suffer from high sparsity (few items per class), while Price distributions can be noisy. Consequently, while they provide supplementary signals, they are less effective than Category/Genre in capturing the primary user preference.
\end{itemize}

\subsection{Case Study}

\begin{figure}[t]
\centering
     \includegraphics[width=0.95\linewidth]{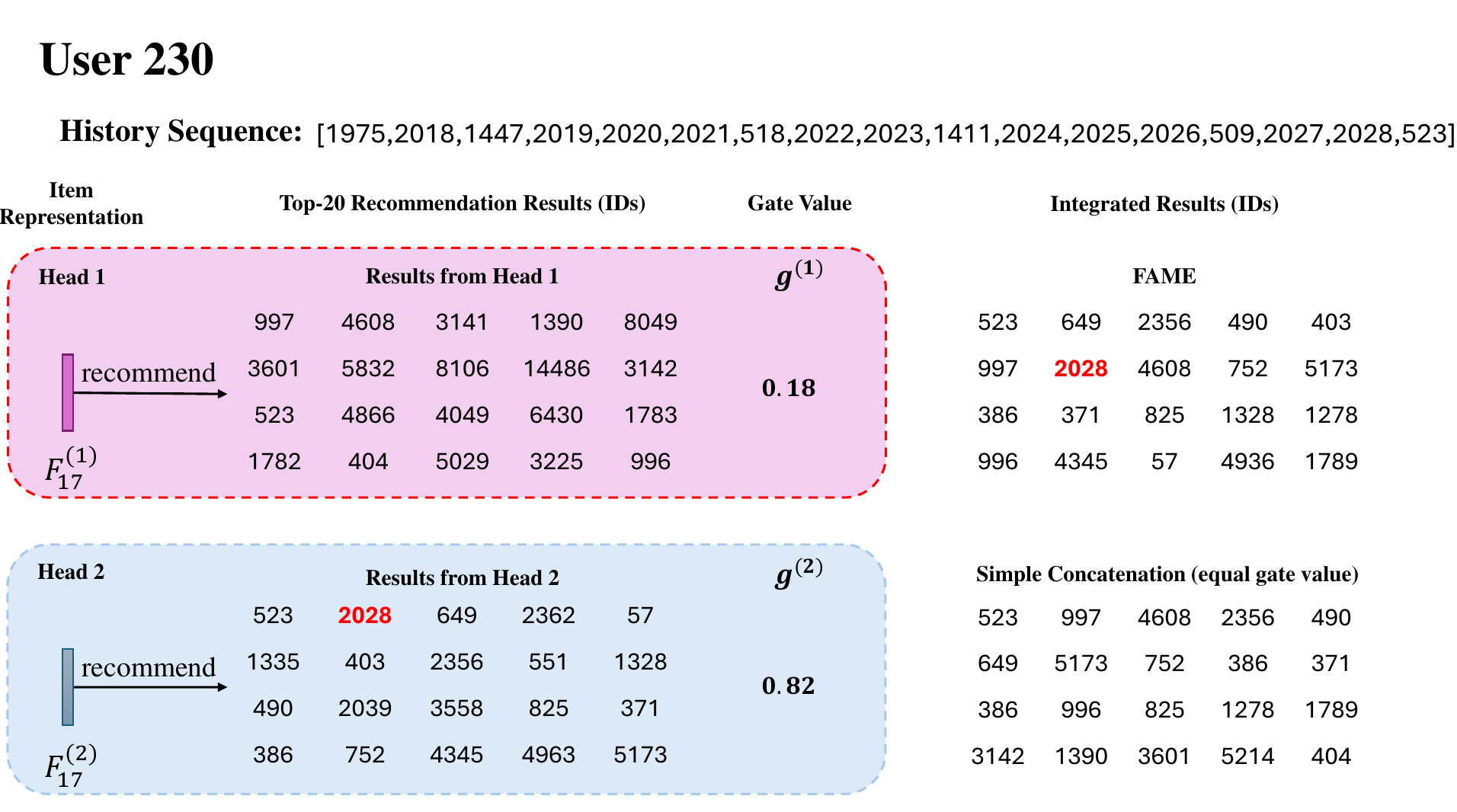}
\caption{Recommendation results for user 230 in the Sports dataset. User history is displayed at the top. The ground truth next item (item 2028) is highlighted.}
\label{fig: case study}
\end{figure}

To illustrate the effectiveness of our facet-aware mechanism, Figure~\ref{fig: case study} presents recommendation results for user 230, along with corresponding head importance scores (calculated using Equation~\ref{eq: gate}). For simplicity, we set the number of heads to two and focus on the Sports dataset. 

The figure clearly demonstrates the diversity of recommendations across different heads, highlighting the ability of our model to capture distinct item facets. The calculated head importance scores reveal that head 2 better aligns with user 230's preferences (0.82 vs 0.18), as evidenced by the inclusion of the ground truth item (item 2028) in its recommendation list. The integrated recommendation, incorporating both heads with appropriate weights, successfully predicts the ground truth item.

In contrast, a traditional approach concatenating sub-embeddings from all heads without considering head importance fails to capture the user's dominant preference, resulting in the omission of the ground truth item in the recommendation list.



\section{Conclusion}
In this work, we proposed \textbf{\textit{FAME}} (\textit{Facet-Aware Multi-Head Mixture-of-Experts Model}), a novel sequential recommendation framework designed to capture the complex, multi-faceted nature of user intents. By repurposing the multi-head attention mechanism, FAME utilizes sub-embeddings from distinct heads to predict the next item independently, effectively modeling diverse item facets (e.g., Genre vs. Director) without increasing model complexity. Furthermore, we integrated a Mixture-of-Experts (MoE) network within each attention head to disentangle fine-grained user preferences within each facet, utilizing a dynamic router network to adaptively aggregate expert outputs based on context.
To address the limitations of random ID-based initialization, we extended this framework to \textbf{FAME+} by introducing a \textit{Text-Enhanced Facet-Aware Pre-training} module. This module leverages pre-trained language models to extract rich semantic features from item metadata and employs an alternating supervised contrastive learning objective. This strategy explicitly disentangles semantic features into facet-specific subspaces before sequential training begins, ensuring robust initialization.
Extensive experiments on four public datasets demonstrate that FAME consistently outperforms state-of-the-art baselines. Notably, the text-enhanced FAME+ variant yields significant performance gains, particularly in sparse data scenarios, confirming that explicit semantic alignment effectively mitigates the cold-start problem and enhances the model's ability to capture multifaceted user preferences.


\bibliographystyle{ACM-Reference-Format}
\bibliography{sample-base}


\end{document}